\documentclass[12pt,preprint]{aastex}

\slugcomment{Accepted by Astrophysical Journal}

\shorttitle{Far-infrared Observations of L1521F}
\shortauthors{Terebey et al.}

\begin{document}


\title{Far-infrared Observations of the Very Low-Luminosity Embedded Source L1521F-IRS in the Taurus Star-Forming Region}


\author{Susan Terebey\altaffilmark{1},  Michel Fich\altaffilmark{2}, 
Alberto Noriega-Crespo\altaffilmark{3}, Deborah L. Padgett\altaffilmark{3}, 
Misato Fukagawa\altaffilmark{4}, Marc Audard\altaffilmark{5}, Tim Brooke\altaffilmark{3}, Sean Carey\altaffilmark{3}, Neal J. Evans II\altaffilmark{6}, Manuel Guedel\altaffilmark{7}, Dean Hines\altaffilmark{8}, Tracy Huard\altaffilmark{9}, Gillian R. Knapp\altaffilmark{10}, Caer-Eve McCabe\altaffilmark{3},  Francois Menard\altaffilmark{11}, Jean-Louis Monin\altaffilmark{11}, Luisa Rebull\altaffilmark{3}
and the Taurus Spitzer Legacy Team}

\altaffiltext{1} {Department of Physics and Astronomy PS315, 5151 State Univ. Dr., California State 
University at Los Angeles, Los Angeles, CA 90032, sterebe@calstatela.edu}

\altaffiltext{2} {Department of Physics and Astronomy, University of Waterloo, Waterloo, Ontario CAN N2L 3G1, fich@uwaterloo.ca}

\altaffiltext{3} {Spitzer Science Center, MC 220-6, California Institute of Technology, 
Pasadena CA 91125}

\altaffiltext{4} {Japan Aerospace Exploration Agency, Institute of Space and Astronautical Science, Japan}

\altaffiltext{5} {ISDC, Switzerland and Observatoire de Gen\`eve, Switzerland }

\altaffiltext{6} {U. Texas, Austin TX}

\altaffiltext{7} {Paul Scherrer Inst., Switzerland}

\altaffiltext{8} {Space Science Inst., Boulder CO}

\altaffiltext{9} {U. Maryland, College Park MD}

\altaffiltext{10} {Princeton U., NJ}

\altaffiltext{11} {LAOG, Grenoble, France}

\begin{abstract}
We investigate the environment of the very low-luminosity object L1521F~IRS using data from the Taurus Spitzer Legacy Survey. The MIPS 160~\micron\ image shows both extended emission from the Taurus cloud as well as emission from multiple cold cores over a $1\arcdeg \times 2\arcdeg$ region. Analysis shows that the cloud dust temperature is $14.2 \pm 0.4$ K and the extinction ratio is $A_{160}/A_K = 0.010 \pm 0.001$ up to $A_V \sim 4$ mag.  We find $\kappa_{160} = 0.23 \pm 0.046~cm^2~g^{-1}$ for the specific opacity of the gas-dust mixture. Therefore, for dust in the Taurus cloud we find the 160~\micron\ opacity is significantly higher than that measured for the diffuse ISM, but not too different from dense cores, even at modest extinction values. Furthermore, the 160~\micron\ image shows features that do not appear in the IRAS 100~\micron\  image. We identify six regions as cold cores, i.e. colder than 14.2 K, all of which have counterparts in extinction maps or C$^{18}$O maps. Three of the six cores contain embedded YSOs, which demonstrates the cores are sites of current star formation. 

We compare the effects of L1521F~IRS on its natal core and find there is no evidence for dust heating at 160 or 100~\micron\ by the embedded source. From the infrared luminosity $L_{TIR} = 0.024~L_{\sun}$ we find $L_{bol\_int} = 0.034 - 0.046~L_{\sun}$, thus confirming the source's low-luminosity.  Comparison of L1521F~IRS with theoretical simulations for the very early phases of star formation appears to rule out the first core collapse phase. The evolutionary state appears similar to or younger than the class 0 phase, and the estimated mass is likely to be substellar.

\end{abstract}


\section{Introduction}

Investigation of starless cloud cores provides a way to study the initial conditions of star formation, where prestellar cores are those judged to be closest to gravitational cloud collapse. The L1521F cloud core in the Taurus molecular cloud at 140 pc distance  (\citet{tor07} and references therein) is a very dense ($n \sim 10^6~cm^{-3}$) and cold ($T \sim 9~K$) cloud core containing $2 - 5 M_{\sun}$ of gas detected in NH$_{3}$ and other dense gas tracers \citep{cod97,oni99,cra04,shi04}.

\citet{oni99} singled out L1521F (their MC 27) as special in mapping surveys of Taurus cores, and argued that the high density and infall asymmetry seen in the $HCO^+(3-2)$ line indicated a core in the earliest stages of gravitational collapse with a free-fall time scale of $10^{3}$ to $10^{4}~yr$. Consistent with the young age, IRAS observations indicated an upper limit of about $0.1~L_{\sun}$ for the luminosity of any possible embedded object. The presence of infall asymmetries in other tracers and an advanced chemical clock led other authors to similar conclusions \citep{cra04,lee04,soh07}. Models of the cloud structure were consistent with the picture of a dense cold core heated externally by the interstellar radiation field \citep{oni99,cra04}.

The Spitzer Space Telescope \citep{wer04} detected a very low-luminosity infrared source in the L1521F prestellar core, a source too faint to appear in the IRAS catalog  \citep{ter05, bou06}.  \citet{bou06} present submillimeter and independent Spitzer data for the core. The source is detected at all IRAC and MIPS wavelengths ranging from 3.6 to 70~\micron\ in wavelength. In this paper we further present 160~\micron\ data for the cloud. The source's low-luminosity suggests the object is substellar, and its similarity to class 0 sources suggests it is extremely young.  An object that predates Spitzer, IRAM~04191+1522, is a low-luminosity $\sim 0.1~L_{\sun}$ class 0 source in Taurus detected by ISO longward of 60 microns \citep{and99,dun06}, that is now suggested as a very low-luminosity object (VeLLO) \citep{dif07,dun08}. However, IRAM~04191+1522 differs from L1521F by exhibiting both a molecular outflow and centimeter radio emission. Other comparable objects detected by Spitzer include VeLLOs in the L1014 and L328 molecular clouds \citep{you04,lee05, hua06}. The spectral energy distributions (SED) are similar, steeply rising with strong emission at 24 and 70~\micron\ in the infrared. \citet{dun08} present numerous VeLLO candidates from the Spitzer c2d survey. The low-luminosity of the VeLLOs suggests their masses are quite small. The presence of dense cores and infall asymmetries towards the better studied sources suggests some may represent a very early stage of star-formation, of accretion onto a substellar mass central object.

Models of early cloud collapse define two distinct phases, the first protostellar core, when gas in the central region first becomes opaque at $T  \sim 100$ K and heats up to $T  \sim 2000$ K at which point begins the second core phase, when molecular hydrogen dissociates and a hydrostatic object at stellar density forms. The properties of the first core depend sensitively on the initial conditions and assumed physics but are roughly $0.01 - 0.08~M_{\sun}$ for mass, $10~AU$ for radius and $L < 0.1~L_{\sun}$ for luminosity \citep{bos95,bat98,sai08}. Numerical simulations suggest the model luminosity then shoots up to $1.0~L_{\sun}$ as the second protostellar core undergoes runaway collapse of a $0.001~M_{\sun}$, i.e. Jupiter mass object. The second core continues to accrete mass but other properties, such as luminosity, at times less than $\sim$10,000 yr are still unknown, until the object becomes older and more massive, and is detectable as a protostar or young brown dwarf.  \citet{sta05} explore whether this phase would be observable, in the context of collapse models with high infall rates. Prior to collapse, the SED of the model cloud peaks near $190~\micron$ due to heating by the interstellar radiation field. They find significant additional heating occurs after the accretion luminosity exceeds $5.0~L_{\sun}$, producing a peak wavelength as short as $90~\micron$ during the class 0 phase. In the context of these models, it appears VeLLOs are unlikely to modify the submillimeter SED of the cloud.  \citet{you04} reach a similar conclusion for the L1014 VeLLO source. Far-infrared wavelengths also problematic: previous work by \citet{bou88} using IRAS data found that the externally heated molecular cloud contributes a significant amount of $0.6 L_{\sun}/M_{\sun}$ to the infrared luminosity.

To better understand the properties of L1521F~IRS and its relation to early phases of star formation we analyze the MIPS~$160~\micron$ and IRAS~$100~\micron$ data near the source. The molecular cloud appears to dominate the emission at these wavelengths, and our goal is to better define the peak of the SED by determining the contribution of the embedded source to the flux. Using these data we also determine the properties of the L1521F cloud core at 160~\micron\ and relate it to other 160~\micron\ cores in the region.  



\section{Observations}

\subsection{Spitzer Observations}
The Spitzer data near L1521F derive from the Taurus Spitzer Legacy Survey \citep{pad09}, a large $44$ square degree map of the Taurus  star-forming region at seven wavelengths ranging from 3.6 to 160~\micron. The nominal spatial resolution of Spitzer is about 1.7\arcsec\ at 3.6, 4.5, 5.8, and 8.0~\micron\ with the IRAC instrument, and is 6, 18, and 40\arcsec\ at 24, 70, and 160~\micron\  with the MIPS instrument \citep{faz04,rie04}. The region near L1521F was observed on 2005-Feb-23 by IRAC and 2005-Mar-02 by MIPS. See \citet{pad09} for an overview of the survey parameters and data processing.

The 160 ~\micron\ data obtained in fast scan mode mapping does not have enough redundancy to fill in completely all the gaps due to a dead readout and the intermediate gap between the array detector rows (MIPS  Data Handbook V3.3 8/24/2007). Furthermore with only 5  pointings per pixel, the effects of hard radiation hits and saturation translates into small regions without data. To deal with these gaps and to preserve as much as possible the diffuse emission, a  3 pixel by 3 pixel (15 \arcsec\  per pixel) median filter is applied to the image. The net effect is a slight redistribution of surface brightness ($\sim 15\%$) and smearing of the original beam from 40\arcsec\  to about 1\arcmin\ in size. To quantify the effect we measured the resulting spatial resolution in two ways. First we determined a FWHM of 52\arcsec\ for a theoretical 160~\micron\ PRF that was processed using the same 3 pixel window filter. We note that most objects appear extended in the 160~\micron\ mosaic. However we were able to  measure an empirical average FWHM of 59\arcsec\ for five pointlike sources, which we take to be the effective spatial resolution of our 160~\micron\ image.

Table \ref{tbl-source} gives the Spitzer position of the embedded infrared source, measured at 24 \micron\ to be least affected by extinction and hence most accurately detect the central source. Positions determined at shorter IRAC wavelengths are the same within the measurement uncertainties. The source is extended at IRAC wavelengths, hence the fluxes given in Table \ref{tbl-flux}  depend on both the aperture and background determination. The  flux density values are comparable to but differ from those in \citet{bou06}. Given that the source is extended we ascribe the difference as due to a difference in the photometric method.

\subsection{IRAS Observations}

Images at 60 and 100~\micron\ were obtained \footnote{http://irsa.ipac.caltech.edu/data/IGA/} from the High-Resolution IRAS Galaxy Atlas (IGA), an atlas that includes the galactic plane, Orion, Taurus-Auriga, and $\rho$ Oph regions \citep{cao97}. The coordinates of IGA images outside the galactic plane are given in 1950 equatorial coordinates. In order to successfully precess and work with the images, several changes were made to the original FITS headers using the Goddard IDL astron library. First, a misspelled keyword RADECSYS was removed and replaced with the keyword RADESYS. Second, the keyword LONPOLE=0.0 was added. In addition, the reference pixels CRPIXn and reference positions CRVALn were changed to be at the image centers  from their original off-image locations. This last crucial change appeared to minimize differences in various FITS software implementations of the CAR projection type.

The IGA images were generated using the HIRES resolution enhancement algorithm and therefore have higher spatial resolution ($2\arcmin$) than the native IRAS ($3.8\arcmin \times 5.4\arcmin$), ISSA (4.3\arcmin), or IRIS (4.3\arcmin) survey images at 100 \micron\  wavelength \citep{cao97,mil05}. We constructed an empirical point spread function (PSF) using several sources in the Taurus survey region to find an effective spatial resolution at 100~\micron\ of $140\arcsec\  \times 110\arcsec\ $ (FWHM), about a factor of two times lower than that for the Spitzer MIPS 160~\micron\ data. 

\section{Results}

\subsection{Morphology of far-infrared emission}

In Figure 1, far-infrared images of dust emission near L1521F  IRS show a striking similarity in the diffuse emission, indicating that the 160 and 100~\micron\ wavelengths are mostly tracing the same dust component in the Taurus cloud. The images are centered on the position of L1521F IRS, but throughout the 1.1\arcdeg\ x 2.2\arcdeg\ field the emission at 160~\micron\  is extended with respect to the spatial resolution of 59\arcsec\ = 8300 AU = 0.04 pc at 140 pc distance.  

In the central 5\arcmin\ the L1521F cloud core is systematically brighter at 160~\micron\ than at 100~\micron, a signature that the core is colder and denser than the surrounding cloud. In addition, the young stellar objects IRAS 04248+2612 and DG Tau appear as prominent point sources in the lower right corner and indicate regions of higher than average dust temperature. The ÒringingÓ at 100~\micron\ around the two point sources is an artifact of the HIRES processing.  Other Taurus YSOs fall within the region shown (Fig. 1), but only the two named sources show significant emission at these wavelengths.

Previous analyses of the diffuse emission seen by IRAS at 60 and 100~\micron\ show the cirrus can be modeled as emission from dust associated with both the atomic and molecular components of the interstellar medium \citep{ter86,bou88,woo94,abe94}. This extended component is variously referred to as galactic emission, infrared cirrus, or the diffuse interstellar medium.  \citet{bou88a} model the diffuse IRAS cirrus as due to dust heated by the interstellar radiation field and find the dust is not in thermal equilibrium except at the longest (100~\micron) wavelength. The morphology of the diffuse emission in the new MIPS 160~\micron\ data shows it to be fundamentally the same as the IRAS 100~\micron\ emission, albeit viewed at higher spatial resolution.

However, emission at the shorter 60~\micron\ wavelength necessarily implies the presence of warmer dust. This has long been understood as a consequence of the Planck function, where emission from cooler dust shifts to the Wien part of the spectral curve and becomes quite faint. Figure 2 shows extended emission along with numerous point sources in a 60~\micron\ IRAS image; the 70~\micron\ MIPS image (not shown) shows the point sources at higher ($\sim$20\arcsec) spatial resolution but has less sensitivity to the extended component.  The morphology of the 60~\micron\ diffuse emission in Figure 2 appears related to the 160 ~\micron\  emission (Fig. 1) but not cospatial, as if representing emission from the surface layers of externally irradiated clouds. The embedded source L1521F IRS  appears as a marginally detected point-like source in the center of the image. However, the object is found in the IRAS Point Source Rejects Catalog (i.e. detected but not reliable) as R04255+2644, which would explain why \citet{oni99} reported the source was not detected by IRAS.  Apart from the faint point source, there is no extended emission in the central 5\arcmin\ of the IRAS image (confirmed by the 70~\micron\ MIPS image) comparable to that seen at 100 and 160~\micron\ (Fig. 1)  from the cloud core. Our focus in this work is the dense cold component and we do not further consider the 60~\micron\ diffuse emission.

\subsection{Analysis of extended dust emission}

Before determining the temperature and optical depth we first give expressions for the intensity.
Assuming dust  that is in thermal equilibrium at temperature $T$,  then at frequency ${\nu}$ the intensity is given by $ I_{\nu}  =  (1 - e^{\tau_{\nu}}) B_{\nu}(T)$ where $\tau_{\nu}$ is the optical depth and $B_{\nu}(T)$ is the Planck function. In the limit of small optical depth, adequate for our purpose, the intensity becomes
\begin{equation}
 I_{\nu}  \approx  \tau_{\nu} B_{\nu}(T).
\end{equation}
From this expression we see that taking the ratio of the intensity at two frequencies leads directly to the dust temperature, if the frequency/wavelength dependent properties of the dust are known.
In practice a range of dust temperatures will be present, but at a given frequency a limited range of temperatures contributes significantly to the intensity, as one `loses' cold dust whose emission falls on the Wien part of the blackbody curve. The contribution of small stochastically heated grains can be neglected at these wavelengths; equilibrium thermal emission from big grains ($> 0.25 ~\micron$) dominates the emission at  $\lambda \ga 70 - 80 ~\micron$ \citep{li01}.

The optical depth can be expressed in terms of the specific opacity $\kappa_{\nu}$ in units of $cm^2 g^{-1}$,  the density of the gas-dust mixture $\rho$ , and the path length L by,
\begin{eqnarray}
 \tau_{\nu}  & = & \kappa_{\nu} \rho L \nonumber\\
 & = & \kappa_{\nu} \mu m_H N_H, \nonumber\\
 & = & \sigma_{\nu}  N_H,
\end{eqnarray}
and where $\mu = 1.38$ is assumed for the mean molecular weight per hydrogen atom, $m_H$ is the mass of a hydrogen atom,  $N_H$ is the hydrogen column density, and $ \sigma_{\nu} $ is the opacity in units of  $cm^2$ per H atom. The frequency dependence of the opacity is often expressed in terms of the power-law index $\beta$ where $\kappa_{\nu} = \kappa_{o} (\nu/\nu_o)^{\beta}$ at some fiducial frequency $\nu_o$, and where $\beta$ ranges from roughly 1.5 - 2 at infrared wavelengths. Based on their definitions, the relation between extinction and optical depth is given by $A_{\lambda} = 2.5log(e) \tau_{\lambda} =1.0857 \tau_{\lambda}  $ and moreover the extinction opacity $\kappa_{\nu} =  \kappa_{\nu sca} +  \kappa_{\nu abs} $ includes both scattering and absorption. The scattering efficiency is specified by the albedo; at long wavelengths the albedo is zero, hence $\kappa_{\nu} = \kappa_{\nu abs} $ for no scattering, which is the regime that  Eqn (1) describes. 

\subsection{Dust temperature}

Figure 3 shows there is a strong linear correlation between the 160 and 100~\micron\  images. This indicates the emission can be well modeled using a single dust temperature but varying optical depth. The two regions immediately surrounding the sources IRAS 04248+2612 and DG Tau depart from the linear trend, which suggests there is internal heating by these YSOs.  Table \ref{tbl-temp} gives the results of a linear fit to the data, after masking out the warm dust in a $6.5\arcmin\ \times\ 6.5\arcmin\ $ square region near the two named YSOs. In addition, the 160~\micron\  image was smoothed to the approximate 100~\micron\ resolution via convolution by the empirical IRAS 100~\micron\  PSF, and the resulting images sampled at 7 pixels = 56\arcsec, the approximate Nyquist frequency of the IRAS image.  Figure 3 also shows evidence for a slowly varying background component, such as due to zodiacal or galactic emission, that appears as a nonzero intercept. As discussed in section 3.1, the origin of the extended emission has been extensively investigated for IRAS data. This background, which appears as a constant offset in our small field, is not obviously related to the Taurus cloud so we do not attempt to model it here. Our temperature determination, which is based on the slope, does not require us to make any assumptions about the background level.

We find a dust temperature of $14.2 \pm 0.4$ K for this part of the Taurus cloud (region shown in Fig. 1), based on the slope of the fit (Fig. 3) and assuming  $\beta = 2$  for the opacity power-law index. Color corrections were derived using the source emissivity from Eqn (1) and are modest ($\sim10\%$) for both MIPS and IRAS data at this temperature. The value given for $\beta = 1.5$ indicates the uncertainty in the temperature if  $\beta$ differs from the nominal value. An analysis using the completely independent COBE data  \citep{lag98} shows excellent agreement; their Table 2 gives $Td = 13.9 \pm 0.2$ K at a position in common with our Taurus image.

\subsection{Optical depth and opacity at 160~\micron}

To the extent that the dust temperature is constant, then the structure seen at 160~\micron\  (Fig. 1) represents variations in optical depth, i.e. variations in column density. Using Eqn (1) and assuming a dust temperature of 14.2 K, then a constant factor $6.18 \times 10^{-5}$ converts from intensity in $MJy~sr^{-1}$ to 160~\micron\ optical depth (after background subtraction). Figure 4 shows the resulting $\tau_{160}$ optical depth map. Near the embedded YSOs, DG Tau and IRAS 04248+2612, the dust temperature is higher than assumed so the optical depth will be overestimated within 3\arcmin\  of the embedded sources.  The optical depth can be underestimated in cold dense cores such as L1521F, a point that we return to.

In order to convert optical depth to column density (or mass) requires the opacity; however the long wavelength opacity is uncertain in molecular clouds, due to possible grain growth, ice mantle formation, and other factors \citep{oss94,sto95,eva01}. The situation is much better constrained for the diffuse ISM, where \citet{lag99} find $\sigma =  8.7 \pm 0.9 ~ 10^{-26} (\lambda / 250~\micron)^{-2} cm^2~H^{-1}$ using COBE data, which appears compatible with \citet{wei01} dust models. Scaling to the MIPS effective  wavelength of 155.9 \micron\  then the diffuse ISM opacity is $2.3~10^{-25}  cm^2~H^{-1}$ from COBE data, which is close to the value of $2.0~10^{-25}  cm^2~H^{-1}$ for $R_V = 3.1$ dust from \citet{wei01,dra03}, and references therein. In alternate units (using $\mu = 1.38$), the gas-dust opacity becomes  $0.098 ~cm^2~g^{-1}$ and  $0.089 ~cm^2~g^{-1}$, respectively. The opacity of dust-only is higher by a factor of the gas to dust ratio,  the value of which is Mgas/Mdust= 124  if using the models of  \citet{wei01,dra03}.

To test whether the diffuse ISM opacity applies in Taurus we compare the 160~\micron\ extinction with extinction $A_V$, which gives an independent measure of the column density. There is a strong correspondence (Fig. 5) between images of  $A_V$ at 4\arcmin\ resolution from 2MASS \citep{fro07} and MIPS $A_{160}$, smoothed to the same resolution. The trend seen in Figure 6 is linear, although it exhibits more scatter than is seen in Figure 3 for the dust temperature. The value of the slope (Table \ref{tbl-dust}) is $A_{160}/A_V = 0.0011$ which is 2.6 times higher than $A_{160}/A_V = 0.00042$ for the diffuse ISM \citep{dra03,wei01}.  This comparison is based on dust with $R_V = 3.1$, the kind found in the Taurus cloud \citep{ken94}; however the predicted ratio hardly changes for $R_V = 4.0$ or $5.5$ dust models. Comparing the observed and theoretical ratios we conclude the 160~\micron\ opacity in the Taurus molecular cloud is 2.6 higher than the diffuse ISM, namely $\sigma = 5.2~10^{-25}  ~cm^2~H^{-1}$  and  $\kappa = 0.23 ~cm^2~g^{-1}$ with estimated one sigma error of twenty percent. This indicates differences in Taurus cloud dust from diffuse ISM dust models, even in molecular cloud regions with modest ($A_V$ = 0.4 to 4 mag) extinction. Since the extinction maps of \citet{fro07} are in fact derived from near-infrared data we express our result relative to K band to find  $A_{160}/A_K = 0.010$, using $A_{V}/A_K = 9.29$ from their assumed dust model. \citet{fla07} report $A_{\lambda}/A_K$ values for shorter Spitzer wavelengths. 

The results we present in Table \ref{tbl-dust} are fundamentally based on an empirical comparison of the 160~\micron\ intensity with $A_V$, hence we further assess the reliability of the $A_V$ determination. In their analysis of a nearby region in the Taurus cloud \citet{arc99} find significant differences between four different $A_V$ methods. In that work they present $A_V$ based on spectral types for 95 stars; Figure 7 shows their $A_V$ values versus  2MASS derived $A_V$ from \citet{fro07}. Comparison with a unit slope line shows good correspondence except for an offset, such that the 2MASS extinction is systematically 0.66 mag smaller than the $A_V$ based on spectral type.  The offset seems plausible given the 2MASS derived value represents an average over  4 \arcmin\ resolution compared with the line-of-sight spectral value. We conclude that, apart from a nonzero offset, the  $A_V$ determination is approximately linear over the $A_V$ = 0.4 to 4 mag range.
 
\subsection{Changes in dust temperature}

As shown in Figure 3, the dust temperature near two young stellar objects, IRAS 04248+2612 and DG Tau, differs from the average 14.2 K value. The luminosities of  $0.4$ and $6.4~L_{\sun}$ for IRAS 04248+2612 and DG Tau, respectively \citep{whi04}, further suggest that the dust ``hot spots'' are internally heated. We now consider the temperature near L1521F IRS, a source that is much lower in luminosity. Figure 3 shows that data points in the L1521F core follow the average trend, and hence show no signature of dust heating. Since the heating effect might be small, in Figure 8 we also plot the {\it unsmoothed} 160~\micron\  versus the IRAS 100~\micron\  data from Figure 1, sampled at the 160~\micron\  resolution. The resulting increased structure at 160~\micron\  leads to increased vertical scatter, as is seen. However notice that the points in the L1521F core all lie systematically above the other data points. This suggests a temperature effect leading to extra 160~\micron\ emission from {\it colder}  than average dust. We conclude there is no evidence for dust heating at 100 or 160~\micron\ by the embedded infrared source in L1521F. We turn this statement into a flux density upper limit. Consider the peak emission value for L1521F, namely the highest 160~\micron\  point in Figure 3 for L1521F. If this data point were moved horizontally to the right, increasing its 100 micron intensity by about $3 MJy~sr^{-1}$ then we would plausibly conclude there is evidence for dust heating. We use our empirical 100~\micron\ PSF to convert this intensity increment into an upper limit of 2 Jy for the 100~\micron\ flux density from circumstellar heated dust around L1521F IRS. The situation is different at shorter wavelengths where the emission is dominated by warm circumstellar dust rather than the cold cloud. The emission at 60~\micron\ shows this behavior (cf. Fig 2), where a near point-like source is detected, but not emission from the extended dense core.

Previous studies at longer wavelengths demonstrate the presence of cold dust in L1521F. Moreover \citet{kir05} determine a characteristic value of 10 K from fitting the observed spectral energy distribution for a number of dense cores. Temperatures of some well determined Taurus cores range from 8 to 12 K \citep{kir07}. The models of \citet{eva01} predict varying dust temperature such that the center of the prestellar core is the coldest region. The physical explanation is that the central region is partially shielded from the interstellar radiation field by the outer layers and thus can reach a lower temperature.  Results from \citet{kir07} give $9.3 \pm 0.3~K$ for the dust in the L1521F core.  In addition, \citet{cod97} find a gas temperature of T = 9.1 K in L1521F using ammonia observations. The gas and dust temperatures are similar and plausibly suggest that the gas and dust may be well coupled at the $n \sim 10^6~cm^{-3}$ density found in the L1521F cloud core.

\subsection{Optical depth maps and 160~\micron\ cold cores}

To accurately model the optical depth of dense cores requires knowing the dust temperature, which we know for L1521F but not for other cores since we lack longer wavelength data. Hence, we take a different approach to studying the cold core component. We a priori assume that two different temperature components contribute to the 160~\micron\ emission, namely T = 10 K  cores and T = 14.2 K clouds. We furthermore use the IRAS 100~\micron\ image to make our model of the cloud component. Thus we obtain an excess image that is sensitive to the cold cores by subtracting a suitably scaled IRAS 100~\micron\ image from the MIPS 160~\micron\ image. \citet{abe94} used a similar method in the Taurus cloud, but they use 60 and 100~\micron\  IRAS data, which is sensitive to somewhat warmer dust.

Figure 9 shows the resulting core and cloud images, each converted to optical depth using the assigned temperature. Including 10 K dust means the optical depth is significantly higher ($\times14$) in core regions than in Figure 4. This implies that accurate optical depths for the cold cores depend on accurate dust temperatures. The dust temperature is known for L1521F, but has not been measured for the other cores, which adds significantly to the uncertainty. Despite the caveats, the core excess image is quite useful. In Figure 9 the core image consists of features that appear at 160~\micron\ but not at 100~\micron, while the cloud image looks like the 100~\micron\ image. Due to the ringing artifact, we treated the regions around the two bright YSOs differently. We cut out regions around the two point sources and filled them with median values determined over a larger region. The effective smoothing this procedure introduces means the optical depth near the two bright point sources has higher uncertainty. Moreover, as for Figure 4, optical depths very close to the point sources (3\arcmin) should be discounted, because those regions have warmer than average dust (see Fig. 3).

One immediate use of the images is to identify dense cold cores, that show up more clearly than in Figure 4. The core image shows six distinct features that we identify as dense cold cores. Three out of six cores contain embedded YSOs, the previously described L1521F IRS, DG Tau, and IRAS 04248+2612.  {\it All} six cores have counterparts in the optical extinction map catalog by \citet{dob05}. The three cores containing YSOs were mapped by \citet{oni98} and designated as C$^{18}$O cores. Table \ref{tbl-id} gives the 160~\micron\ cold core positions and cross identifications.

\subsection{Properties of the L1521F core}

To better understand our analysis we compare with previous studies of the L1521F core, starting with the extinction. The position of the infrared source is spatially coincident with the peak of the extended emission at 160~\micron ; various studies show the core exhibits small scale structure so therefore the peak extinction $A_V$ should depend on spatial resolution. The extinction based on 2MASS data \citep{fro07} has the advantage that it covers the entire Taurus cloud. However, the achieved spatial resolution of $\sim$ 4\arcmin\ is relatively low and in high extinction regions the method begins to break down leading to values that are systematically too low. This is borne out for L1521F where $A_V$ is the smallest we discuss:  $A_V$ is 3.7 toward the infrared source. The 1\arcmin\ resolution of our 160~\micron\ image is higher, and for our single dust temperature fit (Fig. 4) we find $A_V$ = 8 at the peak position. The $A_V$ drops to 6 if measured with respect to the image minimum, illustrating the uncertainty if a reasonable choice is made for a nonzero background level. We note that the assumed background level is not strongly dependent on image size, as the minimum in the L1521F image is close to the minimum found for the entire Taurus survey. 

However, our analysis shows L1521F is one of the regions that contains colder than average dust. The relevant result is from Figure 9 and the extinction increases substantially for our two temperature model (Fig. 9) to $A_V = 43$ mag, where the 10 K core contributes 38 mag, and the 14.2 K cloud contributes 5 to the total extinction. With background subtraction the $A_V$ drops to 41 mag, a small change. At  yet higher (15\arcsec) spatial resolution the trend of higher $A_V$ continues: \citet{kir05} find a peak $A_V = 130$ mag based on 850~\micron\ continuum data. In conclusion we find that analysis of the 160~\micron\ emission provides a useful tool for determining $A_V$ in high extinction regions, although very accurate dust temperatures are needed to fully utilize the method .

Next we compute the mass of the L1521F core, and compare with previous studies. The mass M  is given by $M = \rho L A$ where $A$ is the projected area of the cloud core. Recalling that $ \tau_{\nu} = \tau_{\lambda},  \kappa_{\nu}  =\kappa_{\lambda} $ then comparison with Eqn (2) shows the mass is related to the optical depth such that,
\begin{eqnarray}
M  & = & A \tau_{\lambda} /  \kappa_{\lambda} , \nonumber\\
    & = &  0.61~M_{\sun\ } (0.23~cm^2~g^{-1} / \kappa_{\lambda}) (D/{140~pc})^2 (\Omega_{pix}/{8\arcsec \times 8\arcsec} ) \sum_{i} \tau_{160_{i}}, 
\end{eqnarray}
where $\Omega_{pix}$ is the pixel solid angle in square arcseconds and the sum is over pixels within the dense core, and after background subtraction. One source of error in the mass is the opacity, that we previously estimated to have 20\% uncertainty. The opacity models of \citet{oss94} treat dust coagulation and ice mantle formation in dense core environments. In their study of dense cores \citet{eva01} consider dust opacities OH2 and OH5 from \citet{oss94}, where they favor OH5 opacity (dust with thin ice mantles in $n = 10^6~cm^{-3}$ density gas).  Both the OH2 and OH5 model opacities are consistent with our 160 \micron\ opacity; specifically for $M_g/M_d = 124$ the model OH2 or OH5 leads to $0.25 ~cm^2~g^{-1}$ or $0.33 ~cm^2~g^{-1}$, respectively, both falling within $2 \sigma$ of our  $0.23 ~cm^2~g^{-1}$ value.  How to compare our value with submillimeter opacities is less clear, but assuming $\beta = 2$ and a gas to dust ratio of 100 instead of 124 gives $\kappa_{850} = 0.01~cm^2~g^{-1}$ and $\kappa_{1200} = 0.005~cm^2~g^{-1}$, values that are about a factor of two lower than the OH5 opacities but are essentially the same as the opacities adopted by \citet{and96,cra04,kir05}. A second source of error is in finding the background level, that we estimate leads to 20\% uncertainty in these extended objects. A third, and dominant, source of error is the temperature assumed for the cold cores. We adopted 10 K as a representative temperature for cold cores. However  \citet{kir07} give $T = 9.3 \pm 0.3~K$ for the L1521F core; using this lower temperature leads to a factor of 1.9 increase in the optical depth and hence the mass. Presumably the $9.3~K$ figure is more accurate, but we also quote masses using the $10~K$ dust temperature to illustrate the uncertainty that may apply to other cold cores found within our field.

The core mass determined by \citet{cra04} is $5.5~M_{\sun\ }$ from the 1.2 mm dust continuum; our measurement is $ 7.0, 3.7~M_{\sun\ }$ for $9.3~K, 10~K$ dust temperature, respectively. The mass is calculated using 136\arcsec\ radius aperture, minus a background from a slightly larger radius to mimic the on-the-fly mapping process.  Similarly, \citet{kir05} find a core mass of $2.6 ~M_{\sun\ }$ from the 850~\micron\ dust continuum within 75\arcsec\ radius aperture; within the same aperture we find $2.8, 1.5 ~M_{\sun\ }$ for $9.3~K, 10~K$ dust temperature, respectively. Considering the difference in wavelength and techniques the agreement is good. However, in regions such as L1521F it is clear that an accurate dust temperature for the cold core is needed to determine an accurate mass. 

We point out that the same value of $\kappa_{160}$ derived in modest $A_V = 0.4 - 4$ regions seems to work equally well for the L1521F dense core that contains both higher density and extinction. The opacity we determined for the Taurus cloud is higher than in the diffuse ISM,  but we see no evidence for additional changes in the opacity within different molecular cloud environments.

\subsection{Infrared luminosity of L1521F}

Standard methods of determining bolometric luminosity are problematic for low-luminosity sources such as L1521F~IRS because the emission longward of 100~\micron\ is dominated by heating from the interstellar radiation field. To illustrate the difficulty, \citet{bou88} find an infrared luminosity per solar mass of gas of $1.8~L_{\sun}/M_{\sun}$ for the diffuse ISM and $0.6~L_{\sun}/M_{\sun}$ in molecular clouds; the YSO luminosity thus represents an excess above this `background' luminosity.  In the case of L1521F, prior to Spitzer, the source was successfully modeled purely as a externally heated prestellar core \citep{oni99,cra04}, moreover whose emission longward of $100~\micron$ could be fit by a single temperature greybody \citep{kir05}. The usual definition of bolometric luminosity means to integrate luminosity over all wavelengths. Here we explicitly define $L_{bol\_int}$ to mean just the portion of bolometric luminosity due to internal heating sources. To find $L_{bol\_int}$ the approach taken by several authors has been to fit the source SED based on a physical model \citep{dun06,bou06} and in this way  \citet{bou06} find a best fit value of $0.05 \pm 0.02 ~L_{\sun}$  for L1521F~IRS. Although the method is promising, there is a concern that the applied correction to the observed infrared luminosity is large and possibly model dependent. \citet{dun08} further develop and test this technique and suggest that $L_{bol\_int}$ can be accurately estimated using the $70~\micron$ luminosity.

We adopt an alternative approach of reporting infrared luminosities based solely on the Spitzer data, so that sources can be compared with each other in a robust way. We then introduce SED templates to estimate the internal bolometric luminosity.  One immediate problem is that numerical integration of steeply rising SEDs is sensitive to how the boundary is treated. To avoid this difficulty we adopt several extragalactic conventions \citep{san96} in part because galaxy SEDs share many similarities with embedded YSO SEDs. Specifically, $L_{\nu}(\lambda) = 4 \pi D^2  f_{\nu} $ is defined to be the luminosity per unit frequency. The total infrared luminosity is then a weighted sum of $\gamma \nu L_{\nu}$ over specified bands, where $\gamma$ is a coefficient of order unity. Several definitions of the total infrared luminosity, $L_{TIR}$,  have been proposed (see \citet{huy07} and references therein).  We adopt $L_{TIR} = \gamma_1 \nu L_{\nu}(8.0\micron) +\gamma_2 \nu L_{\nu}(24\micron)+\gamma_3 \nu L_{\nu}(70\micron) +\gamma_4 \nu L_{\nu}(160\micron) $ with $\gamma = [1,1,1,1]$ for simplicity, which is very similar to the $\gamma = [0.95,1.15,1,1]$ proposed by \citet{dra07}.  

One remaining task is to estimate the flux density for L1521F at 160~\micron\, since we detected only the cloud but not the embedded source at this wavelength. We do this by generating an SED template for class 0 sources, using for our template the well-known class 0 source L1527~IRS (IRAS 04368+2557) in Taurus. Because  L1527~IRS is an extended source at 160~\micron\ we briefly discuss its flux density measurement. The flux density is 54 Jy using 64\arcsec\ radius aperture with background determined from a just-larger annulus. The choice of aperture and background is important, and to some extent arbitrary, which seems unavoidable. For L1527~IRS  the measured HWHM size at 160~\micron\ is 50\arcsec, meaning the 64\arcsec\ aperture is slightly larger. To compare with KAO measurements of  \citet{lad91} we use the background at $300\arcsec$;  our flux density then increases to 76 Jy (estimated $20\%$ uncertainty), which compares well with the  total flux of $94 \pm 33$ Jy found by \citet{lad91}. In our SED we use the 54 Jy flux density as the value that better represents the circumstellar emission. Table \ref{tbl-lum} contains measured flux density values for L1527~IRS  while Figure 10 shows the SED.

The total infrared luminosity $L_{TIR} = 1.4~L_{\sun}$ is dominated by the 70 and 160~$\micron$ fluxes, and moreover compares favorably with the IRAS result  $L_{IRAS} = 1.6~L_{\sun}$ \citep{ken90}.  Note that the different beam sizes, with IRAS being the larger (5\arcmin), likely account for any differences. The source has been recently modeled by several groups. \citet{fur08} report  $L_{bol\_int} = 1.8~L_{\sun}$ for the star plus disk luminosity from a model fit to the SED. \citet{tob08} present extensive modeling of the SED and IRAC images of this edge-on source. They determine $L_{bol\_int} = 2.0~L_{\sun}$ from integrating an aperture-defined model SED, and furthermore note that not all photons appear in the SED so the internal luminosity is in fact higher, leading to $ 2.6 ~L_{\sun}$ for the total value. In terms of the SED fit, the model undershoots the long wavelength part of the SED, which suggests that external heating by the ISRF is important, even for this relatively luminous YSO. To conclude, the Spitzer luminosity is $L_{TIR} = 1.4~L_{\sun}$ and based on the models, the internal bolometric luminosity ranges from $\sim 1.9~L_{\sun}$ to $ 2.6 ~L_{\sun}$ which implies a multiplicative bolometric correction of 1.4 to 1.9, respectively.

Returning to L1521F~IRS we find $L_{TIR} = 0.024~L_{\sun}$ (Table \ref{tbl-lum}) where we have estimated $\nu L_{\nu}(160~\micron)$ using the SED template. The estimated bolometric luminosity is then $L_{bol\_int} = 0.034 - 0.046~L_{\sun}$, reflecting the uncertainty in the bolometric correction.  This value is consistent with the luminosity of $0.05 \pm 0.02 ~L_{\sun}$  found by the \citet{bou06} model and $L_{int} = 0.03~L_{\sun}$ from \citet{dun08}. Our result confirms the extremely low-luminosity of the source. 

\subsection{Comparison of L1521F~IRS with the class 0 source L1527}

We continue our comparison of L1521F~IRS with L1527 (IRAS 04368+2557), the one and only previously known class 0 that falls within the boundary of the 44 square degree Taurus Spitzer Legacy Survey. Neither source was detected by 2MASS at 2.2 \micron, consistent with their deeply embedded natures. However, the lower extinction at  IRAC wavelengths (3.6 - 8 \micron) and high sensitivity mean that Spitzer is sensitive to the outflow, showing reflection nebulosity with bipolar morphology toward both sources \citep{bou06,tob08}. 

Figure 11 shows the two objects side by side and illustrates the difference in both spatial extent and source inclination. The symmetric morphology  of L1527~IRS implies an edge-on $i \sim 80 - 90\arcdeg\ $  inclination, that is confirmed by an 8 \micron\  dip in the SED (Fig. 10) that is due to deep silicate absorption \citep{fur08,tob08}. By contrast, the morphology of L1521F~IRS in both our data and that of \citet{bou06} finds the source near the apex of a conical, rather than a symmetric bi-conical structure at 3.6, 4.5, and 5.8 \micron; the fact that one outflow lobe is much brighter than the other in IRAC images implies the source has a moderate  inclination \citep{whi03,ter06}. The radiative transfer models of \citet{whi03} and references therein show the importance of inclination on the SED and images. Based on our simulated images of a protostar with an opaque envelope  \citep{ter06} we estimate the inclination range is $i \sim 50 - 70\arcdeg\ $ for L1521F~IRS. The smaller spatial extent as well as the moderate inclination of L1521F~IRS further show that its luminosity is intrinsically low rather than low due to an inclination effect. 

The $1\arcmin$ scale bar in Figure 11 shows the approximate diameter of the infalling region, where the radius is given by $r = at$ for a sound speed $a = 0.2~ km~s^{-1}$ (appropriate for Taurus) and time $ t = 10^5$ yr \citep{ter84,shu87}. The age of L1527 is not well known but recent results from the c2d Spitzer Legacy Survey give  $1.3 \times 10^5$ yr for the class 0 lifetime \citep{eva08}. The scale bar thus shows the approximate size of the collapsing cloud for class 0 objects. If the VeLLO source were the same age, then in the context of low-mass star formation theory the mass of the collapsing region would be about $2 M_{\sun}$ for the same size region (see section 3.7).

\section{Discussion}

\subsection{Evolutionary status of L1521F IRS}

The interest of L1521F~IRS and other VeLLO sources is where they fit into the evolutionary scheme of low-mass star-formation. Theory predicts an extended accretion phase, the embedded phase, during which a protostar grows in mass, and hence, luminosity, and displays the morphology of infalling cloud core (aka infall envelope), circumstellar disk, and outflow \citep{ter84,shu87}. The embedded phase is represented by class 0 and I sources, where class 0's appear younger, based in part on being more deeply embedded but nonetheless having a central component capable of driving an outflow \citep{and93}. 

Although much lower in luminosity, VeLLOs share many similarities to known class 0 sources, including a steeply rising SED and evidence for outflows. For the purpose of comparison we consider VeLLO sources separately from class 0 sources, as their evolutionary status is still being defined. For the VeLLO  L1521F and class 0 L1527, the cloud core masses are similar but the total infrared luminosity of L1521F~IRS ($0.024~L_{\sun}$) is roughly 50 times smaller. The low-luminosity of L1521F~IRS demonstrates it has a low central mass, regardless of whether the luminosity is dominated by $L_{*}$, the central object luminosity, or   $L_{acc} = f_{x} G M_{*} \dot M_{*}/ R_*$, the accretion luminosity. Because the accretion luminosity is proportional to stellar mass, one interpretation for the low-luminosity of L1521F~IRS  is that it may be younger than L1527~IRS provided the mass infall rates are similar. Given the likely importance of initial conditions at this early phase it seems unlikely that the assumption of constant mass infall rate applies, so it would be an overinterpretation of theory to say that L1521F~IRS is 50 times younger. \citet{sta05} predict luminosities above $1~L_{\sun}$ and would thus suggest VeLLO sources are late class I phase. However, the infall rates in \citet{sta05} are high compared with typical values for Taurus and the model duration is a relatively short $1.6 \times ~10^4$ yr to form an $0.5~M_{\sun}$ protostar. Using infall rates appropriate for Taurus we argue that the morphology and luminosity suggest that  L1521F~IRS could be younger or comparable in age to the class 0 L1527, while the similar mass reservoir of the cloud cores suggests that over time L1521F~IRS may eventually accrete enough mass to become a full fledged star. The cloud core has no well-defined edge, but contains at least $5~M_{\sun}$, thus the source tests the idea that the final stellar mass is determined by the available mass reservoir and will approach a relatively constant fraction of the core mass \citep{mot98,oni98,tes98,joh01,alv07,eno08}.

At this point the statistics in Taurus suggest similar numbers of VeLLO and previously known class 0 sources in Taurus. To distinguish the two types we choose a luminosity boundary of $L_{TIR} = 0.5~L_{\sun}$, which corresponds to roughly $L_{bol\_int} = 1.0~L_{\sun}$. To motivate this choice we calculate that $M_{*} = 0.1~M_{\sun}$ leads to an accretion luminosity $L_{acc} =  1.0~L_{\sun}$, assuming a standard infall rate of  $1.6 \times 10^{-6} ~M_{\sun}~yr^{-1}$, of which fraction $ f_{m} = 2/3$  falls onto a protostar of radius $1 ~R_{\sun}$ (Fig. 7 of \citet{sta88}) and is converted to accretion luminosity with $ f_{x} =0.62$ efficiency \citep{shu96}. Further discussion of efficiencies appears in \citet{ter06}. For sources whose luminosity is dominated by the accretion luminosity, the implied mass  $M_{*} = 0.1~M_{\sun}$ is close to the stellar-substellar boundary, so this choice of luminosity boundary distinguishes between objects that are stellar rather than substellar.

In addition to L1521F and L1527, which lie within the Legacy Survey boundaries, there are two objects in the southern part of Taurus, namely the VeLLO IRAM 04191+1522 and the suggested transitional class 0/I source L1551NE (not to be confused with L1551~IRS5)  \citep{and99,dun06,mor95}.  If the star formation rate has been constant in Taurus then equal numbers implies equal time duration for the two types. Admittedly the numbers are so small that conclusions are premature, except to say that there is no evidence in Taurus that VeLLOs represent a short-lived (10$\times$ shorter) phase compared with current time scale estimates for class 0 objects.  The survey results of \citet{dun08} present 15 new VeLLO candidates, so there is the promise of much better statistics in the near future.

Existing theoretical simulations of the first protostellar core phase appear to be incompatible with the observations of L1521F~IRS. This is because the first protostellar core is relatively cold and produces little or no emission shortward of 70~\micron\ \citep{bos95,bat98,sta05,sai08}, whereas L1521F~IRS is well detected at all IRAC wavelengths. The next phase, the formation of the second (hydrostatic) core, produces emission at the requisite IRAC wavelengths. However, the existing simulations use high infall rates, and hence produce accretion luminosity $L > 1~L_{\sun}$, which is many times higher than the luminosity of L1521F~IRS. A lower infall rate is plausible, hence the source L1521F~IRS might represent the beginning of the second protostellar core phase. There is also the possibility that infall rates are variable. If the current infall rate is very low then the accretion luminosity can drop below $L_{*}$ for the central object. Theoretical evolutionary tracks are relevant for this case although still highly uncertain at young ages. The tracks presented in \citet{bar02} suggest that $ 0.1~M_{\sun}$ is the maximum possible mass for $L< 0.1~L_{\sun}$ if the age is less than 1~Myr. From this we conclude the current mass of L1521F~IRS is likely to be substellar.

Returning to Figure 11, we resume our comparison of the VeLLO and class 0 sources. It is clear that L1527 and L1521F-IRS are different now (extent of nebulosity, luminosity, etc). However both sources could, in fact, have the same (substellar) mass and their differences plausibly be explained by very different accretion rates onto the central protostar (e.g. \citet{ter06}). If the accretion of L1527 were ``turned off'', or the accretion onto L1521F~IRS ``ramped up''  then they could have the same SED, luminosity, CO outflow, and spatial extent. If the accretion rate were, for example, time variable then L1527 could represent a high-phase, and L1521F~IRS a low-phase during class 0 evolution.

The data presented here do not offer a conclusive statement about the final mass of the protostar in L1521F~IRS, whether it will remain a substellar object or continue to grow in mass, as the L1527 class 0 source appears likely to do. However, L1521F~IRS is embedded near the center of a collapsing cloud core \citep{oni99,lee04}, and it is likely to continue to accrete material (possibly at a variable rate) from an ample mass reservoir until infall is terminated by one of possibly many physical processes \citep{ter84,shu87,mye08,eva08}. Given its location inside the cloud core, we suggest L1521F~IRS will not remain substellar unless some condition, such as high angular momentum or ejection, sucessfully terminates accretion soon.

\section{Conclusions}

Using data from the Taurus Spitzer Legacy Survey, we analyze the 160~\micron\  far-infrared emission near L1521F to investigate the cloud environment and look for heating by the central object. The L1521F prestellar core contains a very low-luminosity embedded infrared source previously detected by the Spitzer Space Telescope. The source's low-luminosity suggests the object is substellar, and its similarity to class 0 sources suggests it is extremely young. 

We investigate the influence of L1521F IRS on its natal cloud by studying the Spitzer MIPS 160~\micron\  and IRAS 100~\micron\ data in a $1\arcdeg \times 2\arcdeg$ region to look for evidence of dust heating at these wavelengths. In Figures 1, 3, and 5 we present the basic data and show (1) the emission is dominated by a uniform dust temperature of 14.2 K except near two bright YSOs in the field and (2) there is no evidence for any dust heating associated with L1521F IRS,  suggesting that its impact on the dust continuum at $\lambda\ > 100~\micron$ is minimal. Note however that at 70~\micron\  the dust emission is dominated by the pointlike embedded source.

The cloud dust temperature is $14.2 \pm 0.4$ K, based on the observed linear correlation between the MIPS 160~\micron\ and IRAS 100 ~\micron\  data. There is also a linear correlation with extinction up to $A_V \sim 4$, from which we derive $A_{160}/A_K = 0.010 \pm 0.001$ for the extinction ratio and $\kappa_{160} = 0.23 \pm 0.046~cm^2~g^{-1}$ for the specific opacity of the gas-dust mixture. The value of the opacity at 160~\micron\  is consistent with dust models that incorporate dust coagulation and/or ice mantle formation in dense core environments \citep{oss94}. The 160~\micron\ opacity is 2.6 times higher than that measured for the diffuse ISM, but is consistent with predictions for dense cores, which suggests that there are significant differences between dust in the Taurus cloud and the diffuse ISM even at modest extinction values. 

The MIPS 160 ~\micron\  data show evidence for a cold core component that is not visible in the  IRAS 100 ~\micron\ data. The analysis produces an excess map at 160~\micron\ that is sensitive to colder regions, namely cold dense cores. In the subregion studied we find 6 cold cores, all of which have have counterparts in extinction maps or C$^{18}$O maps. By assuming 10 K dust temperature for the cold cores we compute the optical depth at 160~\micron\  for both the cloud (14.2 K) and core (10 K) components. Three of the six cold cores contain embedded YSOs and thus indicate the cold cores are regions of current low-mass star formation.

To measure source luminosity in an alternate way to model fitting, we introduce the Spitzer total infrared luminosity $L_{TIR}$ and make use of SED templates. For L1521F~IRS,  $L_{TIR} = 0.024~L_{\sun}$ and $L_{bol\_int} = 0.034 - 0.046~L_{\sun}$, depending on the bolometric correction. This result confirms the low source luminosity of $L_{bol\_int} = 0.05~L_{\sun}$ determined by \citet{bou06} and $L_{int} = 0.03~L_{\sun}$ found by \citet{dun08}.  In addition, we estimate the source inclination is $50\arcdeg - 70\arcdeg$ based on the morphology of the extended emission in the IRAC images.

To-date the Taurus Spitzer Survey contains the known class 0 source L1527~IRS, as well as the VeLLO source L1521F~IRS. The source L1521F~IRS appears to be a substellar object that is at a similar or possibly younger evolutionary state than the class 0 in L1527. The mass reservoir of roughly $2 - 5~M_\sun$ in the core suggests that L1521F is capable of eventually becoming a solar-mass type star. Comparison of L1521F~IRS with theoretical models for the very early phases of star formation rule out the first core collapse phase. However it may be consistent with the early second collapse phase, defined as accretion onto a very low-mass hydrostatic object. Alternatively, if the infall rate is variable, and currently low, then the implied mass could be higher. However the low source luminosity is difficult to explain unless the source is substellar.

\acknowledgments

This work is based in part on observations carried out by the {\it Spitzer Space Telescope}, which is operated by the Jet Propulsion Laboratory, California Institute of Technology, under NASA contract 1407. S. Terebey warmly thanks the Spitzer Science Center, UCLA Dept of Astronomy, Caltech Dept. of Astronomy, and Harvard-Smithsonian Center for Astrophysics for their hospitality. M.~Audard acknowledges support from a Swiss National Science Foundation Professorship (PP002--110504).

{\it Facilities:} \facility{SSO}, \facility{IPAC (IRAS)}.

\clearpage

\begin{deluxetable}{rrrrr} 
\footnotesize 
\tablecaption{L1521F~IRS Source Parameters.\label{tbl-source}}
\tablewidth{0pt}
\tablehead{  \colhead{Name}          
     & \colhead{D} & \colhead{L$_{TIR}$}  & \colhead{RA}    &   \colhead{Dec}  \\
    & \colhead{(pc)} & \colhead{(L$_{\odot}$)}  & \colhead{(J2000)} & \colhead{(J2000)}  }

\startdata 
L1521F IRS~1        &140  &0.02     &04:28:38.92  &+26:51:36.2    \\ 
 
\enddata


\tablecomments{
Infrared luminosity from Table 5 and as defined in the text. 
Positions at other wavelengths are same as  
24~\micron\ position, within measurement uncertainties.
Position uncertainty at 3.6, 4.5, 5.8 \micron\ is 1.4\arcsec, and at 
8.0~\micron\ is 1.8\arcsec.
}

\end{deluxetable}

\begin{deluxetable}{crrrrrr} 
\footnotesize 
\tablecaption{Aperture corrected photometry of L1521F~IRS.\label{tbl-flux}}
\tablewidth{0pt}
\tablehead{  \colhead{Telescope}          & \colhead{$\lambda$}   & \colhead{PSF$^{a}$}  &
	\colhead{Flux density$^{b}$} &\colhead{$\sigma$} \\         
   	& \colhead{(\micron)}   	& \colhead{(\arcsec)}  
	& \colhead{(mJy)} & 	\colhead{(mJy)}     } 
	
\startdata 
Spitzer       & 3.6   &1.66    &0.20 & 0.04$^{c}$ \\ 
 & 4.5 & 1.72 & 0.35 & 0.07$^{c}$  \\
 & 5.8 & 1.88  & 0.48 & 0.10$^{c}$  \\
 & 8.0 & 1.98 & 0.98 & 0.2 $^{c}$  \\
 & 24 & 5.9  & 24. & 1.2 \\
 & 70 & 16.  & 460. & 90. \\
 IRAS        & 100  & 118.  &$<$2000  & ... \\ 

\enddata

\tablecomments{
$^{a}$ Nominal PSF from Spitzer Observer's Manual (IRAC) or Data Handbook (MIPS). 
$^{b}$ IRAC photometry uses radius 2 pixel = 2.4\arcsec\ aperture and 2-6 pixel sky background. MIPS 24 uses PRF fitting. MIPS 70 uses radius 3 pixel = 12.\arcsec\ aperture. $^{c}$Source is extended so flux depends on aperture and background.}

\end{deluxetable}


\begin{deluxetable}{lllllll} 
\footnotesize 
\tablecaption{Taurus cloud dust temperature based on 160 and 100~\micron\ data.\label{tbl-temp}}
\tablewidth{0pt}
\tablehead{      
      \colhead{$C_0$} & \colhead{$\sigma_0$} & \colhead{$C_1$} & \colhead{$\sigma_1$} & 
      \colhead{$\beta$}       &  \colhead{$C_{1corr}^{a}$}  &     \colhead{$T_{dust}$}  	  \\         
     & &  &  & & & \colhead{K}  }  
    
\startdata 
  7.94 & 0.12 & 0.259 & 0.002  &  2.0 & 0.235 &   14.2    \\ 
   &   &  &  & 1.5 & 0.235&  15.1    \\ 
\enddata

\tablecomments{Coefficients and uncertainty of linear fit  to data I100 = C0 +C1*I160 shown in Figure 3.
$^{a}$ Corrected slope after applying MIPS and IRAS color corrections based on Eqn 1.}

\end{deluxetable}

\begin{deluxetable}{llllll} 
\footnotesize 
\tablecaption{Comparison of 160  \micron\ opacity with A$_V$.\label{tbl-dust}}
\tablewidth{0pt}
\tablehead{      
       \colhead{$I_{160}/A_{V}^{a}$} & \colhead{$A_{160_0}^{a}$}  &   \colhead{$A_{160}/A_{V}^{a}$}  & 
         \colhead{$A_{160}/A_{K}^{a}$}  &  \colhead{$\tau_{160}/N_H^{b}$} &  \colhead{$\kappa_{160}^{b}$}    	  \\         
    \colhead{$MJy~sr^{-1} ~mag^{-1}$}  & \colhead{mag} & & &
       \colhead{$cm^2~ H^{-1}$} & \colhead{$cm^2~ g^{-1}$}    }  
    
\startdata 
18.1  & 0.0018  & 0.0011  & 0.010 &  5.2~10$^{-25}$&   0.23 \\ 
\enddata

\tablecomments{Coefficients of linear fit  to data shown in Figure 5. $^{a}$ Estimated uncertainties are 10\% although formal errors are much smaller. $^{b}$ Estimated 20\% uncertainties. Quantities at 160~\micron\ are derived based on $R_V = 3.1$ dust model at wavelength V and with model parameters
$4.896E-22 = C_{ext}$ extinction cross section ($cm^2/H$), $1.870E-26 = M_{dust}$ per H nucleon (gram/H), $1.236E+02 = M_{gas}/M_{dust}$,  0.6735 = albedo \citep{dra03, wei01}.   }

\end{deluxetable}

\begin{deluxetable}{llrrrrrrrr} 
\footnotesize 
\tablecaption{Infrared luminosity of L1527~IRS and L1521F~IRS.\label{tbl-lum}}
\tablewidth{0pt}
\tablehead{ \colhead{Source} &  \colhead{$\lambda(\micron$) }  
         & \colhead{3.6}  &	\colhead{4.5} &  \colhead{5.8}       &	\colhead{8.0} 
	&\colhead{24}     &	\colhead{70} & \colhead{160}   &\colhead{TIR$^{a}$ }   }  
	
\startdata 
L1527~IRS  & $f_{\nu} (mJy)$ & 1.2     & 6.2  &  11.  &   8.2   &    580.  & 29000.$^{b}$   & 54000.$^{c}$  & ... \\
  & $\nu L_{\nu} (\lambda)(L_{\sun})$   & 0.0006 &  0.002 &    0.003  &  0.002   &  0.043    &  0.73   &   0.59  &1.37 \\
  & $\nu L_{\nu} (\lambda) /L_{TIR}$      &... & ... & ... & 0.0001 & 0.03 &0.53 & 0.43 & 1.0 \\
L1521F~IRS  & $f_{\nu} (mJy)$ & 0.20     & 0.35 &  0.48&   0.98  &    24.  & 460.  & ... & ... \\
  & $\nu L_{\nu} (\lambda)(L_{\sun})$   & 0.0001 &  0.0001 &    0.0001 &  0.0002   &  0.002   &  0.011   &   0.010$^{d}$  &0.024$^{d}$ \\
  & $\nu L_{\nu}(\lambda) /L_{TIR}$      &... & ... & ... & 0.009 & 0.07 &0.49 & 0.43$^{d}$ & 1.0 \\

\enddata


\tablecomments{
Fluxes from Table  \ref{tbl-flux} or \citet{pad09}, unless noted. More L1527 fluxes in \citet{tob08}.
$^{a}$ Total infrared luminosity, see text. 
$^{b}$ Source extended at 70~\micron. Total flux density measured with standard 100\arcsec\ radius aperture, 120\arcsec\ - 140\arcsec\ radius background, 1.13 aperture correction. Estimated $20\%$ flux density error.
$^{c}$ Source extended at 160~\micron\ includes diffuse cloud. Central flux density measured with 64\arcsec\ radius aperture, 64\arcsec\ - 96\arcsec\ radius background, background = 170 $MJy~sr^{-1}$(i.e. extended cloud core), 1.38 aperture correction for $T < 30 K$ sources. Estimated $20\%$ flux density error.
$^{d}$ Estimated value based on $L_{\nu} /L_{TIR} =  0.43$, adopted from L1527.      }

\end{deluxetable}

\begin{deluxetable}{lllllcc} 
\footnotesize 
\tablecaption{Cross Identification of Spitzer 160~\micron\ Cold Cores.\label{tbl-id}}
\tablewidth{0pt}
\tablehead{      
 \colhead{ID} & \colhead{R.A.} & \colhead{Dec.} &       \colhead{$l$} & \colhead{$b$} & \colhead{$A_V$ Clump$^1$} & \colhead{C$^{18}$O Core$^2$}    \\         
&     \colhead{(deg)} & \colhead{(deg)} & \colhead{(deg)} &  \colhead{(deg)} &  \colhead{Name} & \colhead{Number}    }  
    
\startdata 

04:27:03+26:05:50 & 66.76  &  26.10 & 171.84  & -15.68 &  1211~P17 &  21 \\
04:27:56+26:19:20 & 66.98  &  26.32 & 171.81  & -15.38 &  1211~P22 &  22 \\
04:28:06+27:10:50 & 67.03  &  27.18 & 171.17  & -14.78 &  1211~P32 &  ...  \\
04:28:39+26:51:40 & 67.16  &  26.86 & 171.50  & -14.91 &  1211~P16 &  20  \\
04:29:13+26:14:20 & 67.30  &  26.24 & 172.07  & -15.23 &  1211~P24 &  ...  \\
04:29:35+26:57:30 & 67.40  &  26.96 & 171.57  & -14.69 &  1211~P21 &  ...  \\
 
\enddata

\tablecomments{  The 160~\micron\  peak position (J2000) using 15\arcsec\ pixels.  \\
$^1$ Cloud name + clump name from 0.1\arcdeg\ resolution catalog by \citet{dob05}. \\
$^2$ Core name from Table \ref{tbl-source}  in \citet{oni98} and based on data with 0.05\arcdeg\ resolution.
}

\end{deluxetable}

\clearpage



\begin{figure}
\plotone{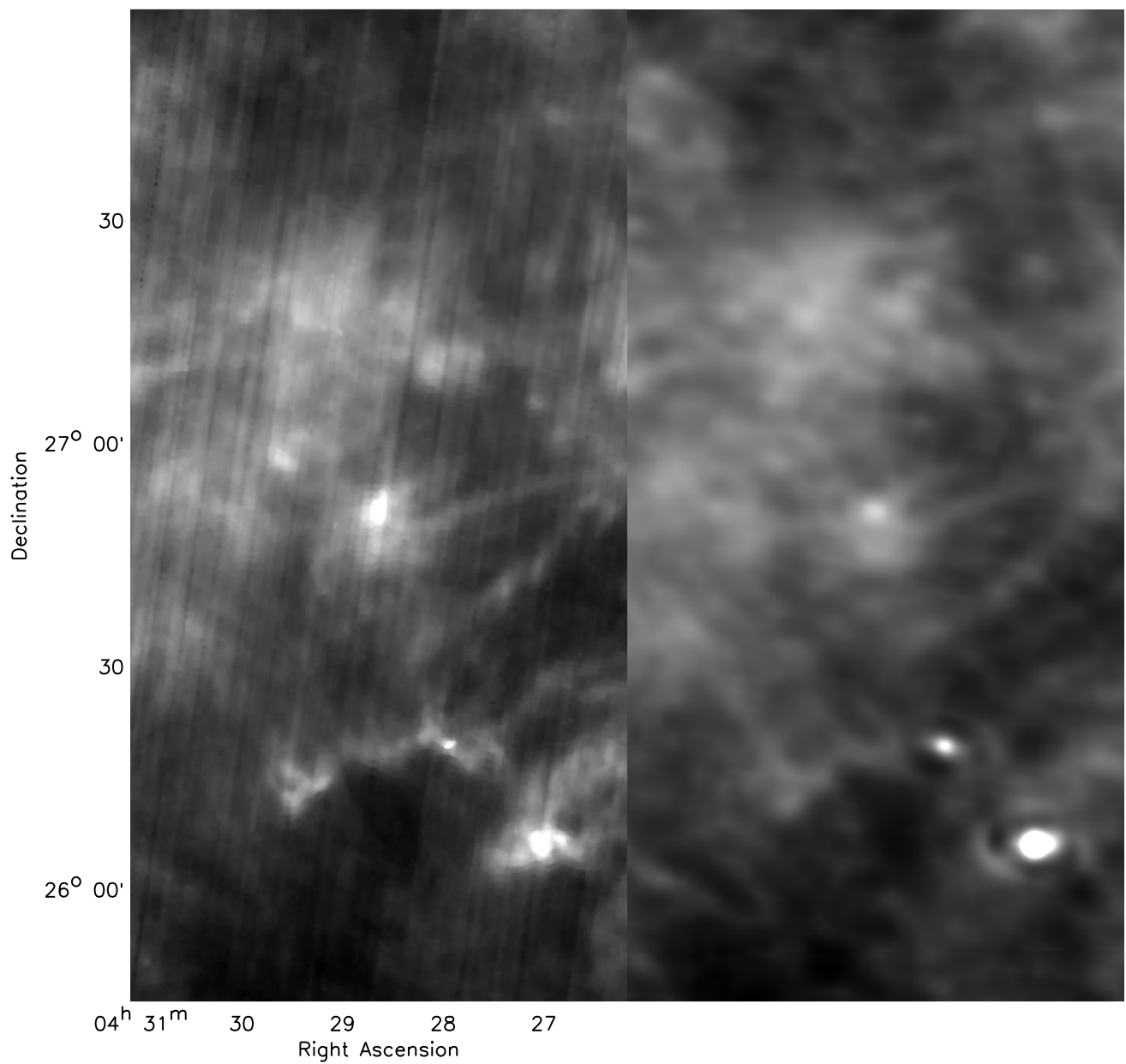}
\caption{Far-infrared images of dust emission in 1.1\arcdeg\ x 2.2\arcdeg\ field centered on L1521F  IRS. The MIPS 160~\micron\ (left panel) and IRAS 100~\micron\ (right panel) images are strikingly similar, indicating the two wavelengths are mostly tracing the same dust component in the Taurus cloud. The L1521F cloud core (at image center) is brighter at 160  \micron\ than at 100~\micron, a signature that the core is colder and denser than the surrounding cloud. Two point sources are prominent in the lower right corner, especially at 100~\micron, and indicate regions of higher than average dust temperature. The two bright sources are IRAS~04248+2612 and DG Tau from E to W.  \label{fig1}}
\end{figure}

\begin{figure}
\epsscale{.50}
\plotone{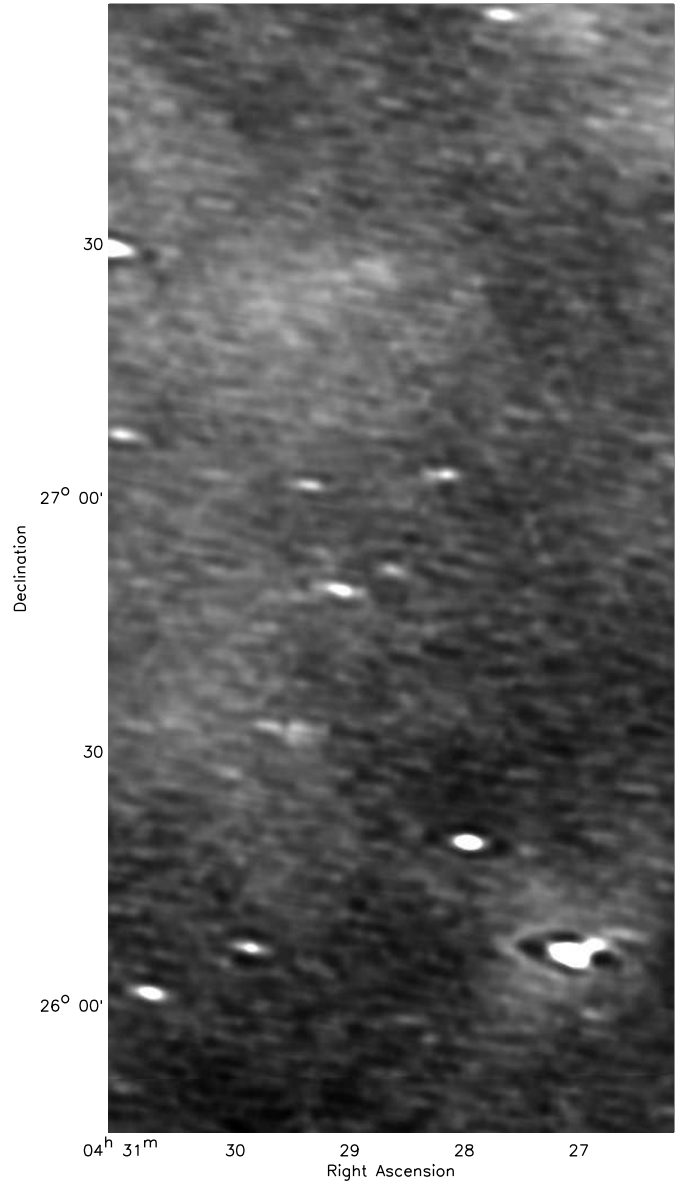}
\epsscale{1.00}
\caption{Extended dust emission is also seen across the IRAS 60~\micron\  image but there is less correlation with the 160~\micron\ image  (Fig. 1). The 60~\micron\ emission appears related but not cospatial, as if tracing warmer dust associated with the cloud surface.  Field of view same as Figure 1. A faint but marginally detected source appears at the the position of L1521F~IRS, exactly at the image center. Most point sources seen at 60~\micron\ do not appear at longer wavelengths.  \label{fig2}}
\end{figure}

\begin{figure}
\plotone{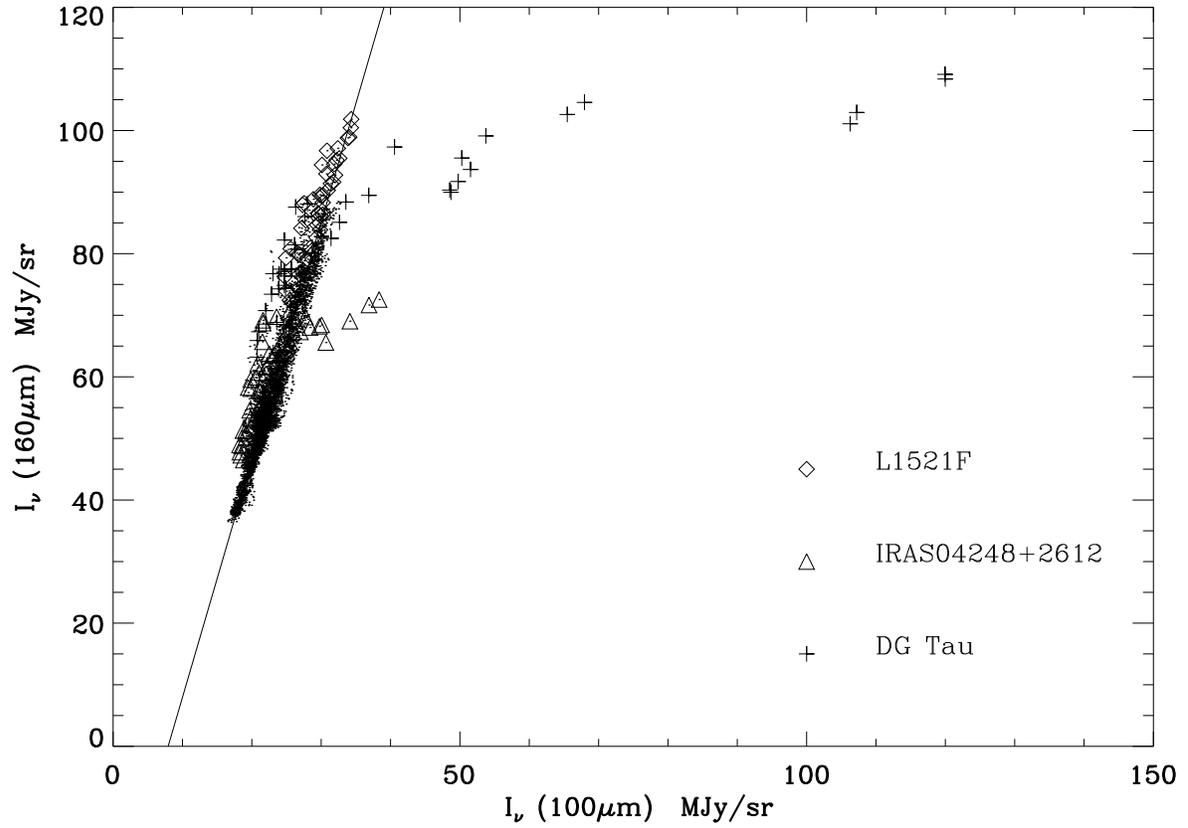}
\caption{There is a strong linear correlation between the resolution matched MIPS 160~\micron\  and  IRAS 100~\micron\ emission. Solid line shows a linear fit to the data while symbols show regions surrounding specific objects. Excess 100~\micron\ emission indicates warm dust around the YSOs IRAS04248+2612 and DG Tau. No evidence of warm dust is seen towards L1521F IRS, suggesting the source luminosity is too low to provide significant dust heating.  \label{fig3}}
\end{figure}

\begin{figure}
\epsscale{.50}
\plotone{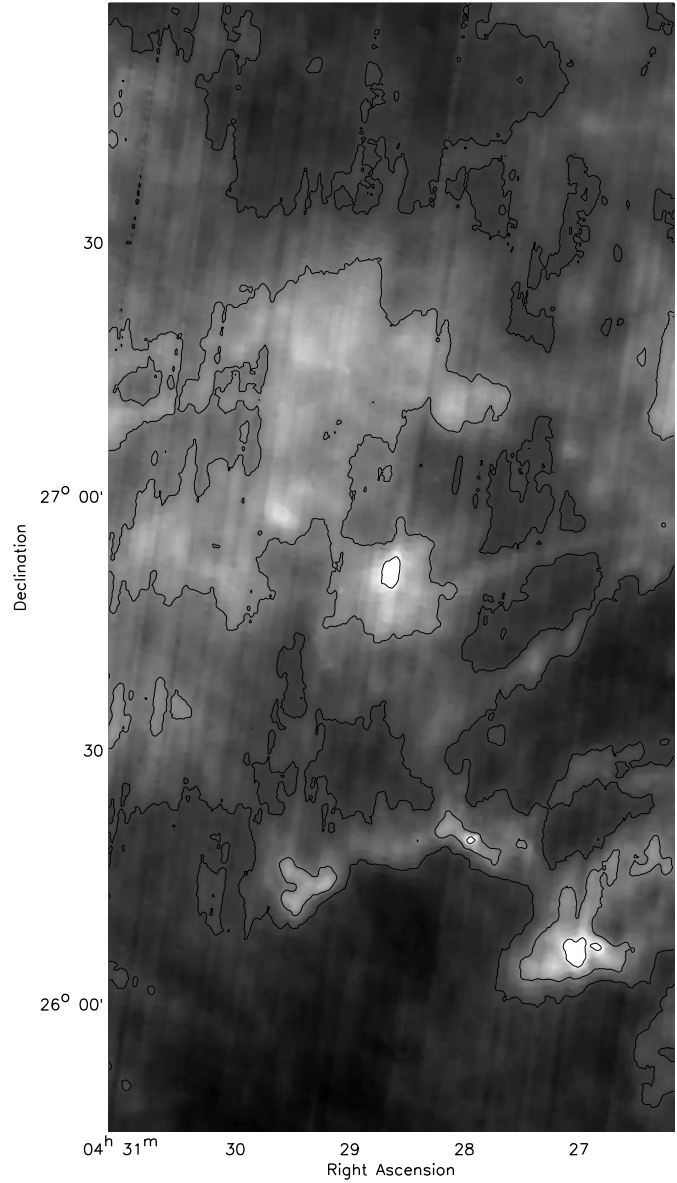}
\caption{Optical depth map at 160~\micron\ using constant dust temperature (using Table \ref{tbl-temp}). Solid lines show 0.0032, 0.0044, 0.008 contour levels. If measured with respect to image minimum (0.002 at 04:29:12.7 25:49:13 position) then the contours correspond to $\delta A_V = 1,2,4$, respectively (using Table \ref{tbl-dust}). Optical depths are less accurate at the highest contour level because the constant temperature assumption breaks down in these regions (see text). \label{fig4}}
\end{figure}

\begin{figure}
\epsscale{1.00}
\plotone{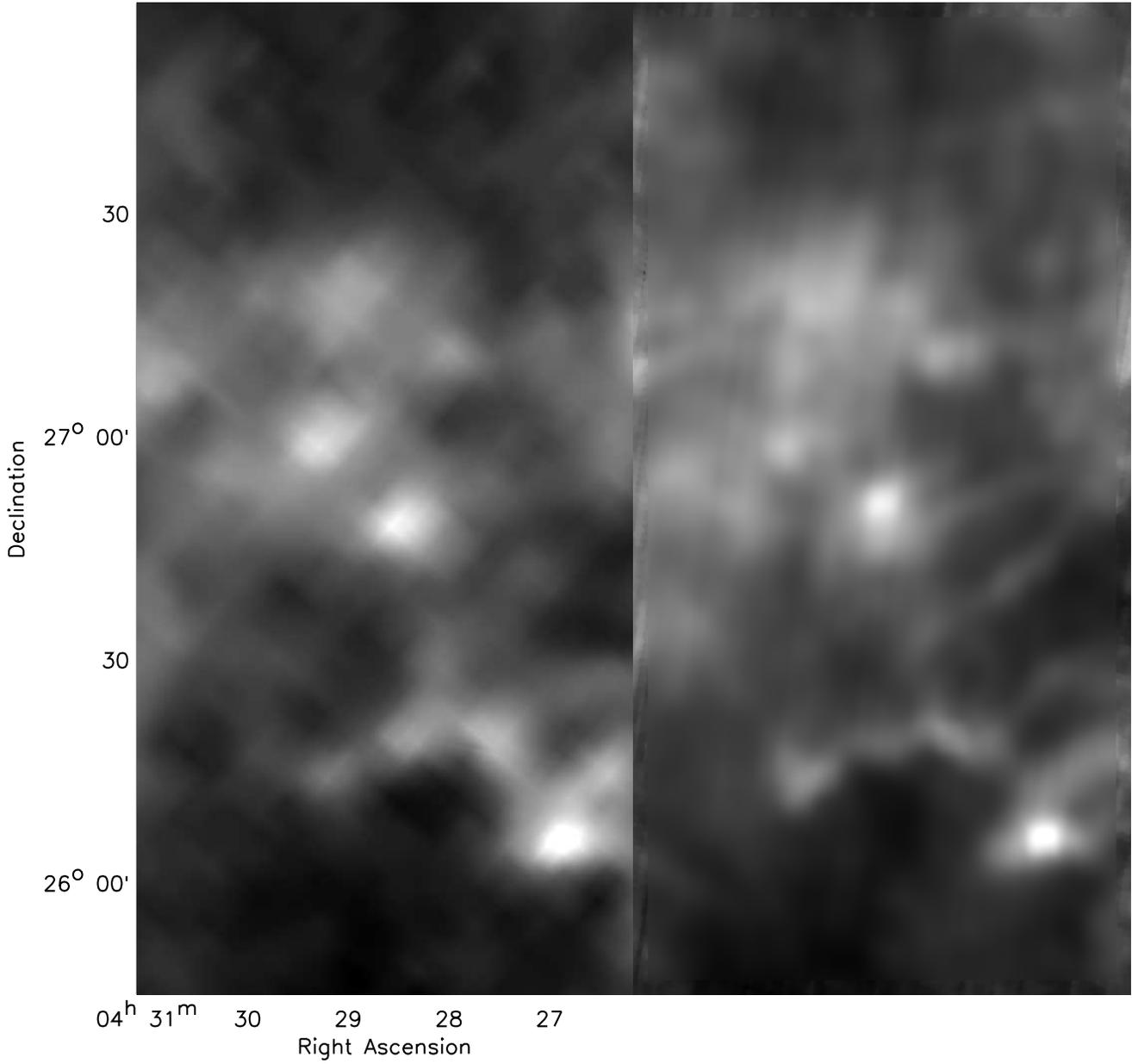}
\caption{Comparison of extinction determined from 2MASS with that from 160~\micron\ data. Left panel is $A_V$ from \citet{fro07}; right panel is extinction $ A_{160}$  (see Fig. 4) but smoothed to have matching 4\arcmin\ resolution.  \label{fig5}}
\end{figure}

\begin{figure}
\plotone{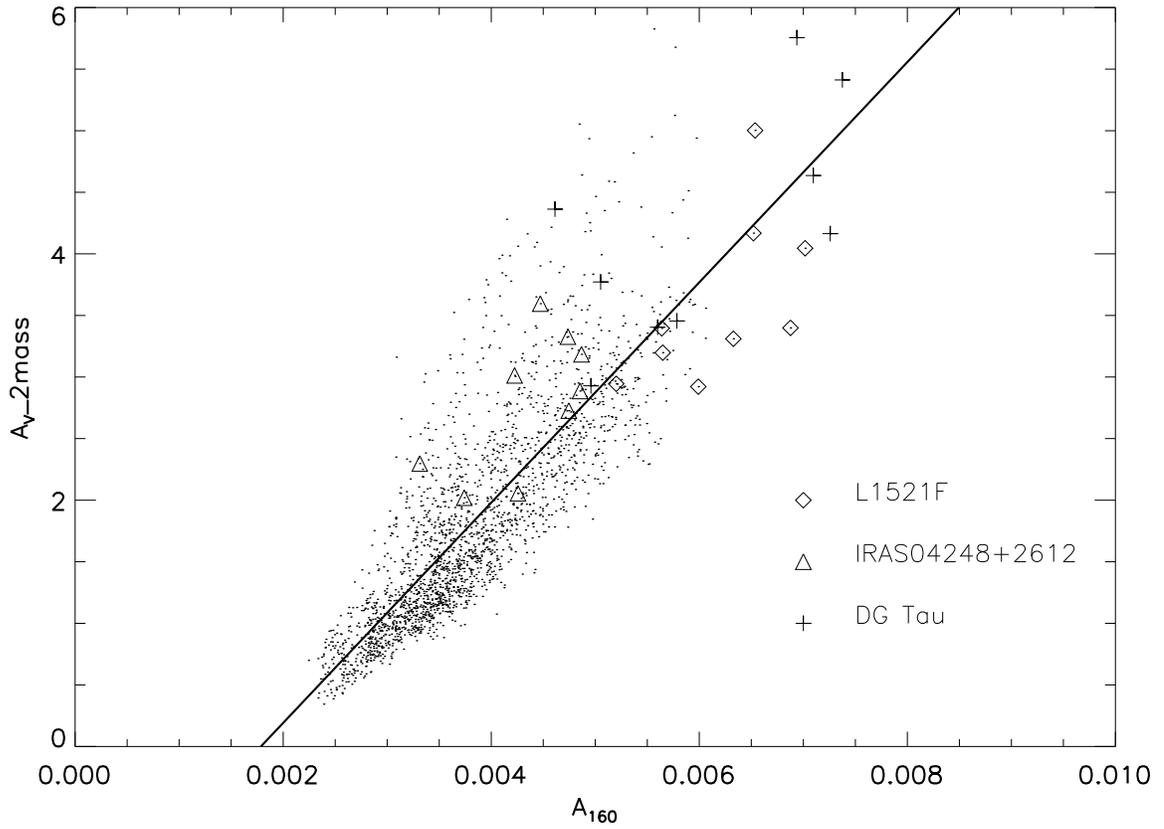}
\caption{There is a linear correlation between  $A_V$ and $ A_{160}$ up to $A_V \sim 4$ for the spatial resolution matched data.  \label{fig6}}
\end{figure}

\begin{figure}
\plotone{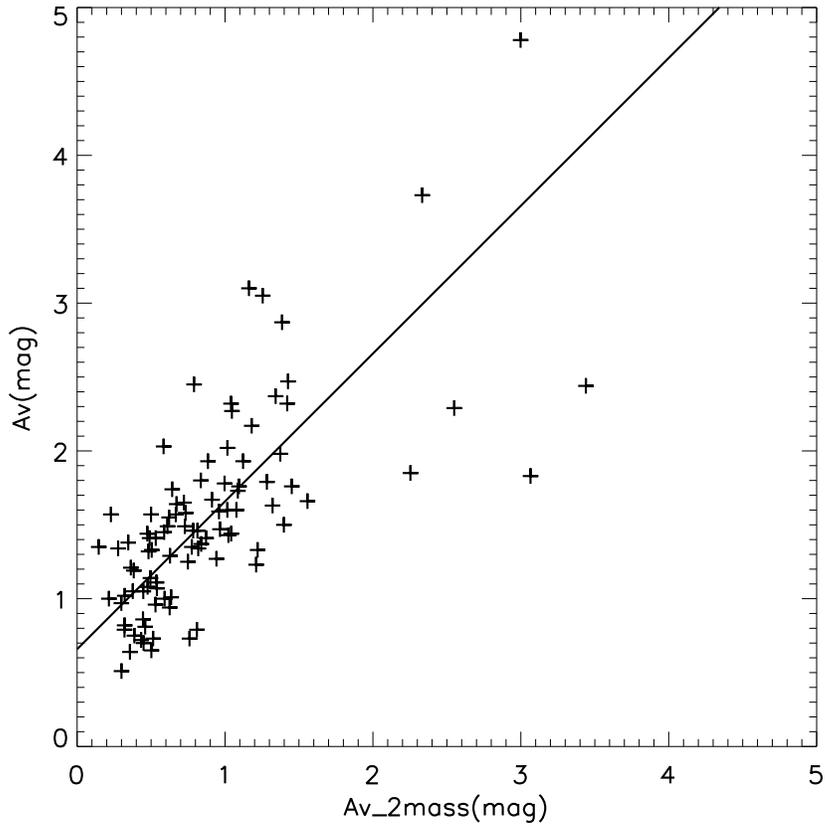}
\caption{Comparison of two different $A_V$ methods shows good agreement with unit slope up to $A_V \sim 4$. However the two methods have a systematic 0.66 mag offset, and above $A_V \sim 2$ the scatter increases.  \label{fig7}}
\end{figure}

\begin{figure}
\plotone{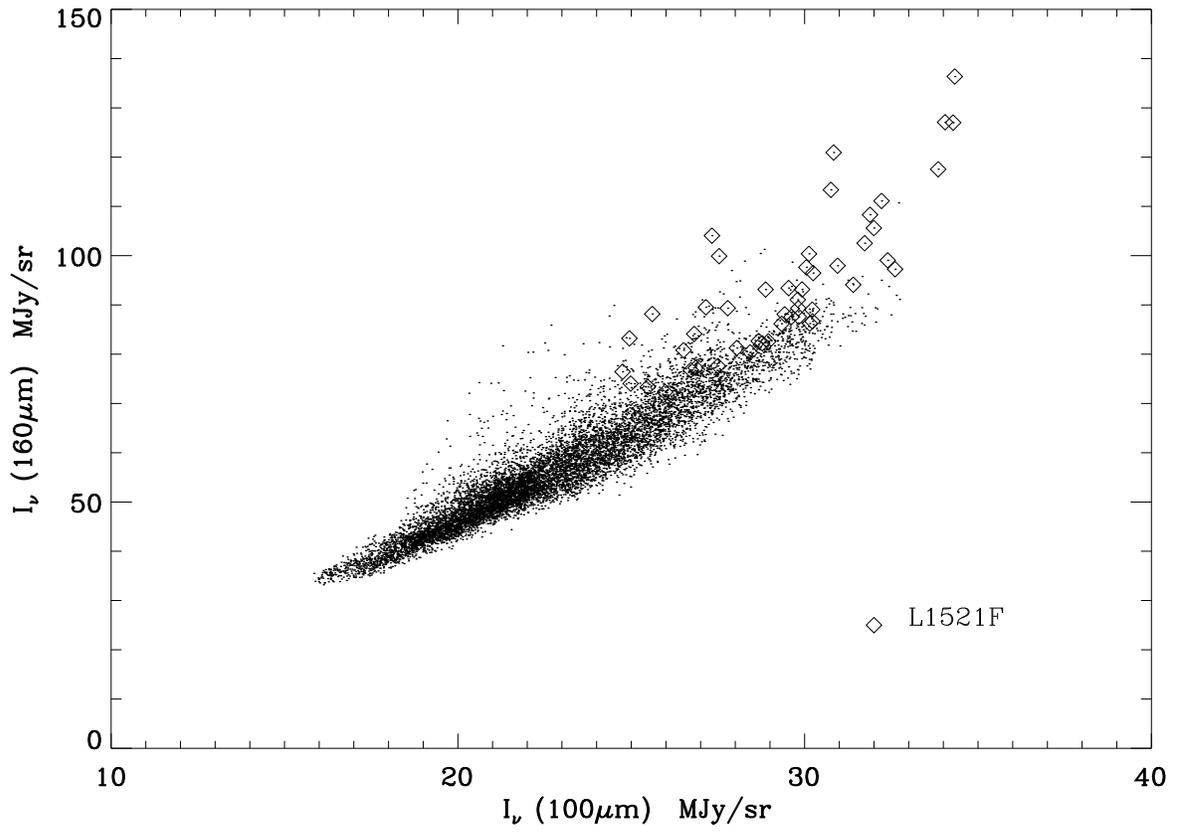}
\caption{Using the unsmoothed 160~\micron\ data more clearly shows the 160~\micron\  excess from the cold dust in the L1521F core. Regions near the two embedded sources have been excluded.   \label{fig8}}
\end{figure}

\begin{figure}
\plotone{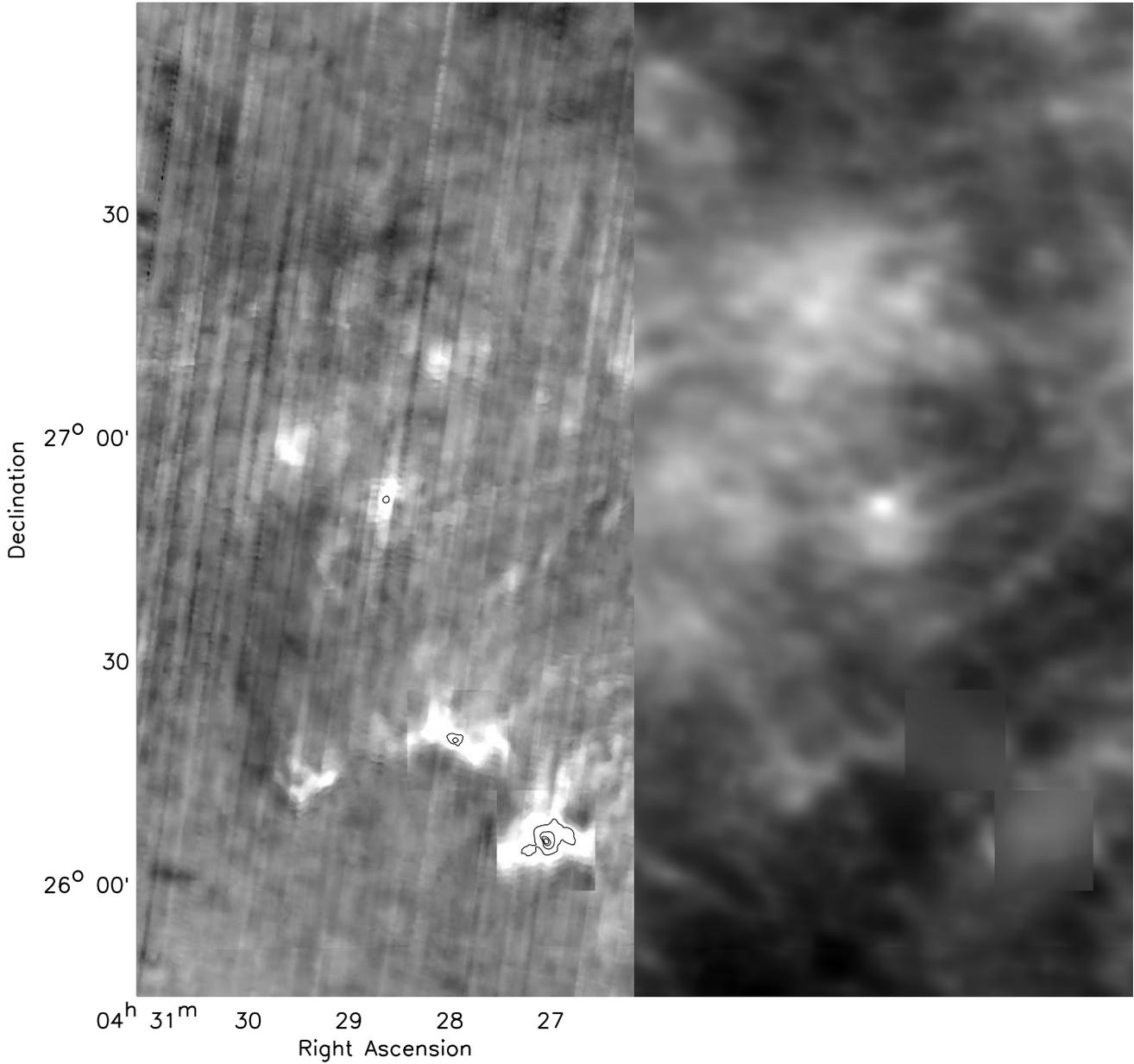}
\caption{Optical depth greyscale maps at 160~\micron\ for core component (left) and cloud component (right).  The 160~\micron\ excess map (left) traces regions that differ from the average behavior (right). The emission has been converted to optical depth for cores (left) assuming $T_D = 10$ K and for the cloud (right) using $T_D = 14.2$ K. Adding the maps gives the total extinction. Optical depth contours start at 0.03 with 0.03 increment.  \label{fig9}}
\end{figure}

\begin{figure}
\plotone{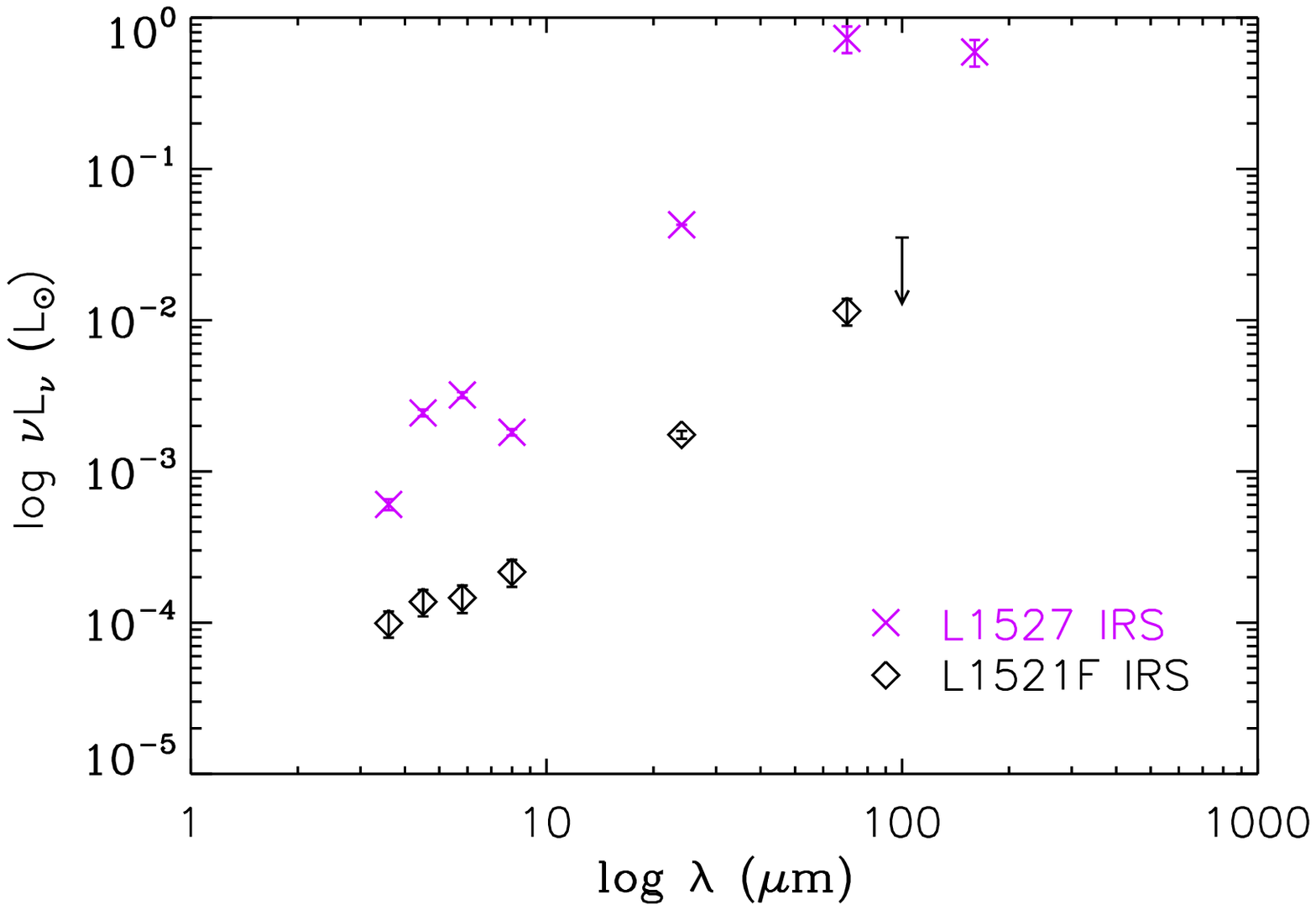}
\caption{Comparison of class 0 and VeLLO  spectral energy distributions. The top SED shows the central point source of L1527~IRS from 3.6  \micron\ to 160~\micron. The dip at 8.0  \micron\  is associated with 10  \micron\ silicate absorption that is due to the edge-on source inclination \citep{fur08,tob08}. The overall SED of L1521F~IRS is lower indicating a lower luminosity. Both sources have steeply rising SEDs showing they are deeply embedded.  \label{fig10}}
\end{figure}

\begin{figure}
\epsscale{1.0}
\plottwo{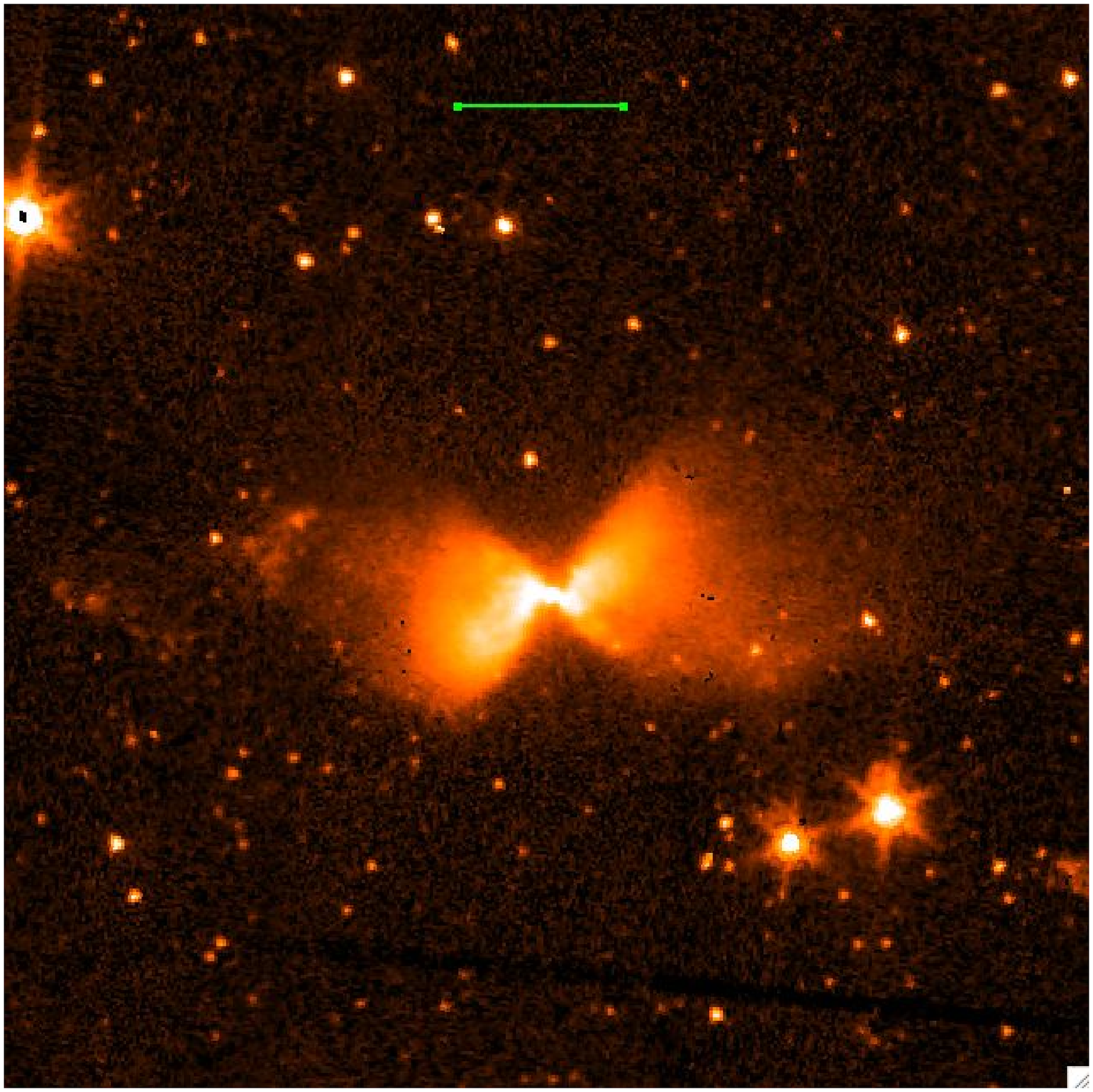}{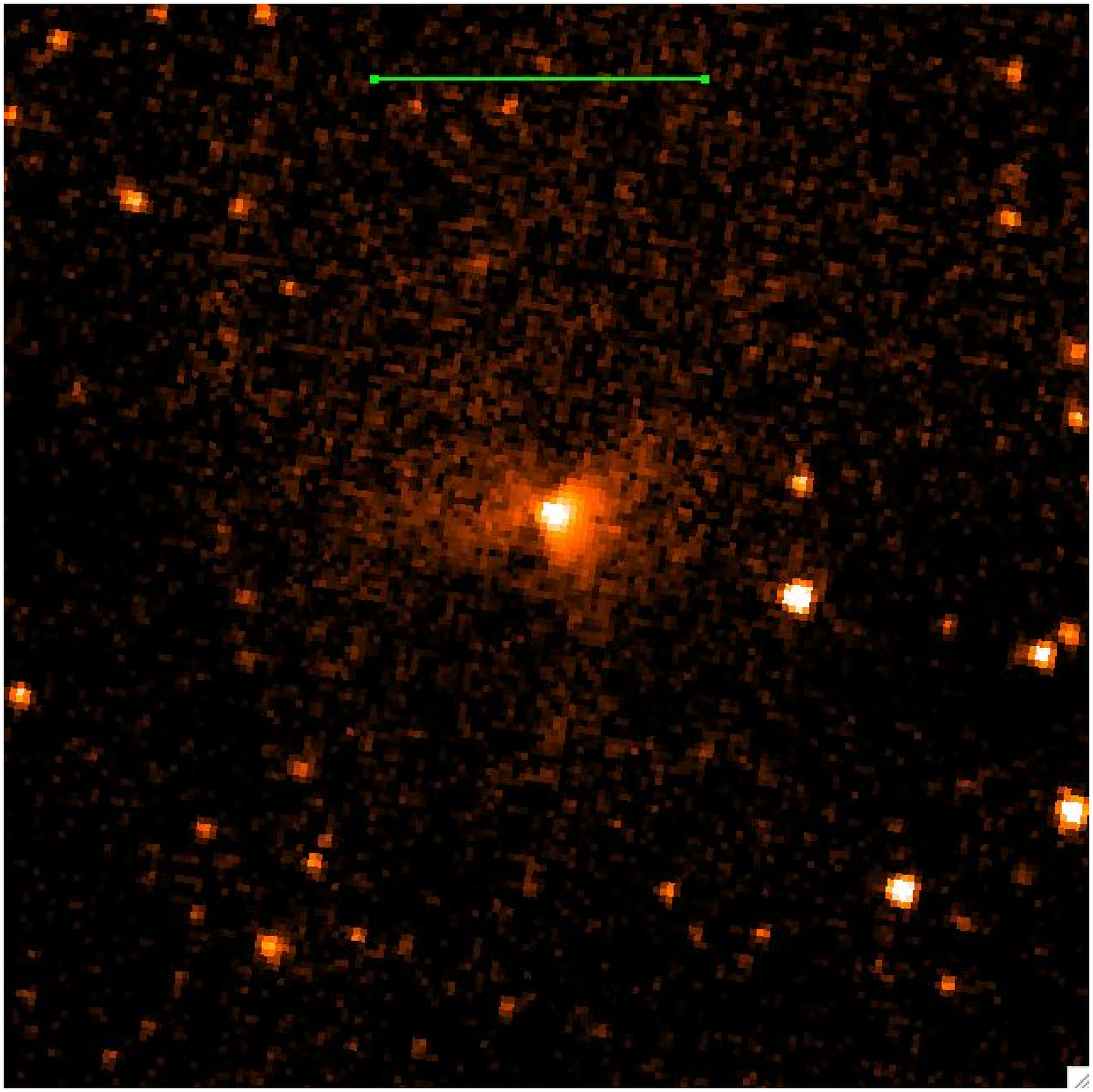}
\caption{Comparison of class 0 and VeLLO images at IRAC 4.5  \micron\ wavelength shows scattered light nebulosity toward both. Scale in green is $1\arcmin$ = 8400 AU. North is up, east to the left. (Left) Image of  L1527~IRS edge-on system shows a disk oriented N-S and outflow cavity E-W in scattered light. (Right)  L1521F~IRS lies at apex of conical nebulosity, suggesting an outflow seen at moderate inclination.  The smaller extent of the VeLLO nebulosity is consistent with its lower luminosity. \label{fig11}}
\end{figure}

\end{document}